\documentclass[a4paper,fleqn,usenatbib,useAMS]{mnras}

\usepackage{graphicx}	
\usepackage{amsmath}	
\usepackage{amssymb}	
\usepackage{multicol}   
\usepackage{bm}		
\usepackage{pdflscape}	
\usepackage[export]{adjustbox}
\usepackage{subfig}
\newcommand{\ct}{\citealt}

\usepackage[T1]{fontenc}
\usepackage{ae,aecompl}

\usepackage{newtxtext,newtxmath}

\title[Grain Size Effect on Chemistry \& Non-ideal MHD]{Effect of Grain Size on Differential Desorption of Volatile Species and on Non-ideal MHD Diffusivity}

\author[B. Zhao et al.]{Bo Zhao$^{1}$\thanks{Contact e-mail: \href{mailto:bo.zhao@mpe.mpg.de}{bo.zhao@mpe.mpg.de}}\thanks{Present address: Giessenbachstr. 1, D-85748, Garching, Germany},
Paola Caselli$^{1}$,
Zhi-Yun Li$^{2}$\\
\\
$^{1}$Max-Planck-Institut f\"ur extraterrestrische Physik (MPE), Garching, Germany, 85748\\
$^{2}$University of Virginia, Astronomy Department, Charlottesville, USA, 22904\\}

\pubyear{2018}

\begin{document}
\label{firstpage}
\pagerange{\pageref{firstpage}--\pageref{lastpage}}
\maketitle

\begin{abstract}
We developed a chemical network for modeling the chemistry and 
non-ideal MHD effects from the collapsing dense molecular clouds 
to protostellar disks. First, we re-formulated the cosmic-ray 
desorption rate by considering the variations of desorption rate 
over the grain size distribution. We find that the differential 
desorption of volatile species is amplified by the grains larger 
than 0.1~$\mu$m, because larger grains are heated to a lower 
temperature by cosmic-rays and hence more sensitive to the 
variations in binding energies. As a result, atomic nitrogen N 
is $\sim$2 orders of magnitude more abundant than CO; N$_2$H$^+$ 
also becomes a few times more abundant than HCO$^+$ due to the 
increased gas-phase N$_2$. However, the changes in ionization 
fraction due to freeze-out and desorption only have minor effects 
on the non-ideal MHD diffusivities. Our chemical network 
confirms that the very small grains (VSGs: below a few 100~$\AA$) 
weakens the efficiency of both ambipolar diffusion and Hall effect. 
In collapsing dense cores, a maximum ambipolar diffusion is 
achieved when truncating the MRN size distribution at 0.1~$\mu$m, 
and for a maximum Hall effect, the truncation occurs at 0.04~$\mu$m. 
We conclude that the grain size distribution is crucial to 
the differential depletion between CO and N$_2$ related molecules, 
as well as to the non-ideal MHD diffusivities in dense cores. 
\end{abstract}




\section{Introduction}
\label{Chap.Intro}

In dense molecular cloud cores that are slightly ionized 
\citep{BerginTafalla2007}, magnetic fields are expected to partially 
decouple from neutral matter through non-ideal MHD effects, including 
ambipolar diffusion (AD), Ohmic dissipation, and Hall effect 
\citep[e.g.][]{Nakano+2002}. The non-ideal MHD effects are essentially 
determined by chemistry and microscopic physical processes 
\citep{OppenheimerDalgarno1974,UmebayashiNakano1990}. In this way, 
chemistry actually has a significant impact on the dynamics and 
evolution of the system, from the collapsing cloud to the planet 
forming disk. However, until recently, chemistry models in this 
context are mostly simplified, with little attention paid to how 
different chemical effects affect the level of non-ideal MHD 
self-consistently.

When solving the chemical network for non-ideal MHD effects, 
existing literature often considers a simplified chemical network 
and treats different molecular and ionic species collectively 
\citep[e.g.,][]{UmebayashiNakano1990,Nishi+1991,KunzMouschovias2009,Marchand+2016,Wurster2016,Zhao+2016}.
In these chemical networks, heavy ions are denoted collectively as $m^+$ 
and molecules collectively as $m$, where $m^+$ and $m$ are usually assumed 
to be HCO$^+$ and CO, respectively. Recently, \citet{Dzyurkevich+2017} 
has developed a more complete chemical network (H-C-O chemistry) 
for non-ideal MHD effects, yet the effect of grain size on the Hall 
diffusivity is not fully revealed. 
Hall effect can dominate over ambipolar diffusion when including grains 
of $\sim$10--20 nanometer size, but the values of both diffusivities are 
not necessarily large enough to allow disk formation. 
Indeed, there exists an optimal grain size ($a_{\rm min} \approx 0.04~\mu$m) 
to achieve the strongest Hall effect, which we will clarify in this work. 

In cold dense cores of molecular clouds, the majority of molecules are 
frozen onto grain surfaces. The large CO freeze-out implies that HCO$^+$ 
will no longer be the main positive charge carrier in dynamically evolved 
starless cores (i.e., prestellar cores), but H3$^+$ and its deuterated 
isotopologues instead \citep{Caselli+2003,Flower+2005,Tassis+2012}. 
Such a change in ion composition in dense cores has not been considered 
self-consistently in previous networks for non-ideal MHD diffusivities 
\citep{KunzMouschovias2009,Marchand+2016,Wurster2016,Zhao+2016,Dzyurkevich+2017}.
We will show that the freeze-out of molecules has limited effect 
on the non-ideal MHD diffusivities, because the fluid conductivities are
dominated by small grains instead of molecular ions; however, it can affect 
somewhat the polarity and strength of the Hall effect in the intermediate 
density range 10$^9$--10$^{11}$~cm$^{-3}$ for certain choices of 
grain size.

Another unsolved puzzle in astrochemistry is the survival of nitrogen 
bearing species in the gas phase to higher densities than those 
at which most carbon bearing species are already depleted out 
\citep{Tafalla+2004,BellocheAndre2004,Bergin+2002}. 
While the difference in binding energies of CO and N$_2$ is only 
about 10\% \citep{BerginLanger1997,Oberg+2005} and hence is not 
considered as the origin of the differential depletion between 
nitrogen and carbon bearing species, it is believed that the loss of 
gaseous CO may account for the increase in abundances of N$_2$H$^+$ 
\citep{Aikawa+2001,Bergin+2002,Jorgensen+2004} and thus NH$_3$ 
\citep{Geppert+2004,Aikawa+2005}. However, it is still a second-order 
effect compared to the direct freeze-out of neutrals 
\citep{BerginTafalla2007}. Alternatively, reducing the sticking 
coefficients of N$_2$ onto the grains by one order of magnitude 
can also enhance the abundances of both NH$_3$ and N$_2$H$^+$ 
\citep{Flower+2005}, yet the approach is somewhat ad hoc. 
In this work, we present a new way of enhancing the gas-phase 
abundances of N, N$_2$ and N$_2$H$^+$ by including the grains size 
distribution into the formulation of cosmic ray desorption rate. 
However, due to the large binding energy for NH$_3$ adopted in this work, 
it is still difficult to maintain a large amount of gas-phase NH$_3$ 
that are efficiently formed via the usual gas-phase route from N$_2$ 
\citep{Caselli+2017}. More accurate measurements on the binding energies 
of molecules, as well as the inclusion of surface chemistry 
are required to better resolve these puzzles. 

The rest of the paper is organized as follows. Section \ref{Chap.Archem} 
describes the set up of the chemical network and the modeling of 
reaction rates. In Section \ref{Chap.Result}, we show the effect 
of grain size distribution on the differential desorption of volatile 
species, followed by a more complete discussion of the effect of grain 
size on non-ideal MHD diffusivities. 
Finally, we summarize our main result in Section \ref{Chap.Summary}.

\section{Chemical Network}
\label{Chap.Archem}

We developed a reduced chemical network (Table~\ref{Tab:specs}) 
including the 21 major neutral species in dense molecular cloud, 
31 corresponding ion species, electron, and neutral and singly charged 
grain species. 
We consider over 500 reactions including gas phase reactions, recombination 
of charged species on grains, as well as freeze-out onto and thermal 
desorption of molecules off grains. In contrast to some existing networks 
\citep[e.g.,][]{Garrod+2008}, 
we include all possible charge transfer reactions 
involving grains (for each size bin; hence the total number of 
reactions depends on the number of bins), which are crucial for 
obtaining correct ion abundances in the high density regimes where 
grains are the dominant charge carriers.
\begin{table}
\caption{Chemical Species}
\label{Tab:specs}
\centering
\begin{tabular}{c}
\hline\hline
{\bf Neutral Species} \\
\hline
H$_2$, H, He, \\
C, CH, CH$_2$, CH$_3$, CH$_4$, \\
N, N$_2$, NH, NH$_2$, NH$_3$, \\
O, O$_2$, OH, H$_2$O, \\
CO, CO$_2$, Mg, Fe \\
\hline\hline
{\bf Ion Species} \\
\hline
H$^+$, H$_2^+$, H$_3^+$, He$^+$, HeH$^+$, \\
C$^+$, CH$^+$, CH$_2^+$, CH$_3^+$, CH$_4^+$, CH$_5^+$, \\
N$^+$, N$_2^+$, N$_2$H$^+$, NH$^+$, NH$_2^+$, NH$_3^+$, NH$_4^+$, \\
O$^+$, O$_2^+$, O$_2$H$^+$, OH$^+$, H$_2$O$^+$, H$_3$O$^+$, \\
CO$^+$, HCO$^+$, CO$_2^+$, HCO$_2^+$, NO$^+$, Mg$^+$, Fe$^+$ \\
\hline\hline
{\bf Electron \& Grain Species} \\
\hline
e$^-$, g$^-$, g$^+$, g$^0$ \\
\hline\hline
\end{tabular}
\end{table}

We use the standard grain size distribution $n(a)\propto a^{-\rm Ind}$ 
with a fixed power law index {\it Ind} between the minimum size 
$a_{\rm min}$ and maximum size $a_{\rm max}$. 
The entire size range is divided logarithmically into 20 size bins. 
The total grain mass is fixed at $q=1\%$ of the gas mass. The density 
of grain material is set as $\rho_g=3.0$~g~cm$^{-3}$. 

\subsection{Reactions}
\label{S.React}

The primary mission of a chemical network is to determine the abundance 
of each species at any given time by solving the coupled rate equations. 
The rate equation for each species $i$ of number density $n_i$ can 
be expressed by the difference in its formation and destruction: 
\begin{equation}
\begin{split}
{{\rm d} n_i \over {\rm d}t} = & \sum_{j,k} k_{jk} n_j n_k + \sum_{l} k_l n_l - n_i \sum_m k_{im} n_m \\ 
& + (k_{\rm des}+k_{\rm crd}+k_{\rm crd,fl}) n_{i,s} - k_{\rm acc} n_i~,
\label{Eq:rateEQs}
\end{split}
\end{equation}
where we include formation of species $i$ by two-body reactions ($k_{jk}$) 
and cosmic-ray ionization ($k_l$), and destruction of species $i$ by 
two-body reactions ($k_{im}$), freeze-out onto dust grains ($k_{\rm acc}$), 
thermal ($k_{\rm des}$) and cosmic-ray desorption from dust grains 
($k_{\rm crd}$), as well as desorption by cosmic-ray induced secondary 
UV photons ($k_{\rm crd,fl}$). 

The gas-phase reaction rates are taken from UMIST database 
(\ct{McElroy+2013}; see also \ct{Tomida+2013} and \ct{Marchand+2016}), 
formulated by the Arrhenius representation, 
\begin{equation}
k(T) = \alpha \left({T \over 300~K}\right)^\beta {\rm exp}\left(-{\gamma \over T}\right)~,
\end{equation}
where $\alpha$ is the pre-exponential factor, $\beta$ characterizes 
the temperature dependence of the rate coefficient, and $\gamma$ is 
the activation energy or energy barrier of the reaction in units of 
degrees K.

We consider only singly charged grains, which is appropriate for 
the grain charge population in dense cores \citep{DraineSutin1987,Ivlev+2015b}.
The reaction rates between gas-phase species and grains are given by 
\citep{KunzMouschovias2009}, 
\begin{eqnarray}
\alpha_{\rm e^- g^0} = \pi a^2 \left({8k_{\rm B}T \over \pi m_{\rm e}}\right)^{1/2}\left[1+\left({\pi e^2 \over 2ak_{\rm B}T}\right)^{1/2}\right]\mathcal{P}_{\rm e}~, \label{Eq:aleg0}\\
\alpha_{\rm i g^0} ~= \pi a^2 \left({8k_{\rm B}T \over \pi m_{\rm i}}\right)^{1/2}\left[1+\left({\pi e^2 \over 2ak_{\rm B}T}\right)^{1/2}\right]\mathcal{P}_{\rm i}~, \label{Eq:alig0}\\
\begin{split}
\alpha_{\rm e^- g^+} = & \pi a^2 \left({8k_{\rm B}T \over \pi m_{\rm e}}\right)^{1/2}\left[1+\left({e^2 \over ak_{\rm B}T}\right)\right] \\ & \times \left[1+\left({2 \over 2+(ak_{\rm B}T/e^2)}\right)^{1/2}\right]\mathcal{P}_{\rm e}~, \label{Eq:aleg+}\\
\end{split}
\\
\begin{split}
\alpha_{\rm i g^-} = & \pi a^2 \left({8k_{\rm B}T \over \pi m_{\rm i}}\right)^{1/2}\left[1+\left({e^2 \over ak_{\rm B}T}\right)\right] \\ & \times \left[1+\left({2 \over 2+(ak_{\rm B}T/e^2)}\right)^{1/2}\right]\mathcal{P}_{\rm i}~; \label{Eq:alig-}\\
\end{split}
\end{eqnarray}
where $k_{\rm B}$ is the Boltzmann constant, $m_{\rm i}$ and $m_{\rm e}$ 
are the ion and electron mass, respectively, and $\mathcal{P}_{\rm i}$ 
and $\mathcal{P}_{\rm e}$ are the sticking probabilities of ions or 
electrons onto grains, whose values are assigned as 1.0 and 0.6, 
respectively \citep{Umebayashi1983,KunzMouschovias2009}.

The rate coefficients for charge transfer between charged grains are 
given by,
\begin{equation}
\begin{split}
\alpha_{\rm g^-g^+} = & \pi a_{\rm sum}^2 \left({8k_{\rm B}T \over \pi m_{\rm red}}\right)^{1/2}\left[1+\left({e^2 \over a_{\rm sum}k_{\rm B}T}\right)\right] \\ & \times \left[1+\left({2 \over 2+(a_{\rm sum}k_{\rm B}T/e^2)}\right)^{1/2}\right]~, \label{Eq:alg-g+}\\
\end{split}
\end{equation}
where $a_{\rm sum}$ is the sum of the radii of two grains, and $m_{\rm red}$ 
($\equiv m_{\rm g^+} m_{\rm g^-} / (m_{\rm g^+}+m_{\rm g^-})$) 
is the reduced mass of the two grains. 

The accretion rate of each species onto dust grains is given by,
\begin{equation}
k_{\rm acc}=\pi a^2 \left({8k_{\rm B}T \over \pi m_{\rm i}}\right)^{1/2} n_{\rm g}(a)~,
\end{equation}
where $n_{\rm g}(a)$ is the total number density of dust grains with size 
$a$. 

The thermal desorption rate of species $i$ is given by 
\begin{equation}
k_{\rm des}=v_0(i) {\rm exp}\left(-E_{\rm des}(i) \over k_{\rm B}T_{\rm d}\right)~,
\end{equation}
where $T_{\rm d}$ is the dust grain temperature that is set to the gas 
temperature for our application, and $v_0(i)$ is the characteristic 
vibration frequency for the absorbed species $i$, given by the following 
relation, 
\begin{equation}
v_0(i)=\sqrt{2n_{\rm s}E_{\rm des}(i) \over \pi^2 m_{\rm i}}~,
\end{equation}
where $n_{\rm s}=1.5\times10^{15}$~cm$^{-2}$ is the number of surface 
sites per cm$^2$ on grains. 

The cosmic-ray particles of few 10$^1$ to few 10$^2$~MeV per nucleon can 
impulsively heat dust grains to a higher temperature T$_{\rm e}$ 
(depending on the cosmic ray energy and grain size). The grain 
subsequently cools through desorption of volatile species, e.g., CO. 
The rate of such cosmic-ray induced desorption can be approximated by 
\citep{HasegawaHerbst1993}:
\begin{equation}
k_{\rm crd}(i,a)=f(T_{\rm e}(a))k_{\rm des}(i,T_{\rm e}(a))~,
\label{Eq:k_crd}
\end{equation}
where $f(T_{\rm e}(a)$) is an estimation of the fraction of time 
spent by a grain with radius $a$ at an elevated temperature 
$T_{\rm e}$, defined as the ratio of the cooling timescale by 
desorption of volatiles to the time interval of successive heating by 
cosmic-rays. The latter is about 10$^6$~yrs for 0.1~$\mu$m grains and 
cosmic-ray ionization rate of $10^{-17}$~s$^{-1}$ \citep{Leger+1985}; 
this timescale is inversely proportional to the cross-section of the 
grain ($\propto a^2$).

The elevated temperature a grain can reach via impulsive heating 
is also sensitive to the grain size. Cosmic-ray particles deposit more 
energy when penetrating through larger grains ($\propto a$), but find 
it difficult to heat up the whole grain to high temperatures due to a 
larger grain volume. We extend the volumic specific heat formula up to 
300~K from Fig. 1 of \citep{Leger+1985}, 
\begin{equation}
\begin{split}
C{\rm v}(T) & = 1.4\times10^3~T^2~~{\rm erg~cm^{-3}~K^{-1}}, T<50~{\rm K}~, \\
& = 2.2\times10^4~T^{1.3}~~{\rm erg~cm^{-3}~K^{-1}}, 50~{\rm K}<T<150~{\rm K}~, \\
& = 4.4\times10^5~T^{0.7}~~{\rm erg~cm^{-3}~K^{-1}}, 150~{\rm K}<T<300~{\rm K}~.
\end{split}
\label{Eq:Cv}
\end{equation}
The energy deposit to a grain with radius $a$ by impinging Fe nuclei 
with $\sim$20~MeV per nucleon is estimated by 
\begin{equation}
\Delta E_{\rm dep}=3.8\times10^5({a \over 0.1 \mu m})~{\rm e}V~,
\label{Eq:E_dep}
\end{equation}
which corresponds to heating up the grains of 0.1~$\mu$m to 70~K and 
0.2~$\mu$m to 43~K. Smaller grains can reach much higher $T_{\rm e}$, 
e.g., $\sim$193~K for 0.03~$\mu$m grains and over 600~K (extrapolated)
for grains $\lesssim$0.01~$\mu$m. 

The subsequent cooling of the grain is mainly achieved through evaporation 
of CO and other volatile species. According to \citet{HasegawaHerbst1993}, 
the evaporation timescale is inversely proportional to $k_{\rm des}$ of 
CO molecules ($\approx$10$^{-5}$~s for 0.1~$\mu$m), which is 
applicable to grains with enough surface sites for volatile absorbates. 
From the initial deposited energy $\Delta E_{\rm dep}$, we can estimate 
the maximum number of CO molecules that would be evaporated into the gas 
as $N_{\rm evap}=\Delta E_{\rm dep}/E_{\rm des}({\rm CO})$. 
$N_{\rm evap}$ is usually a small fraction of the total number of 
surface sites; but for grains smaller than $\sim$0.03~$\mu$m, 
$N_{\rm evap}$$\approx$10$^6$ becomes comparable to the 
number of surface sites (assuming a few ice layers). Therefore, 
we only apply the scaling of evaporation timescale from 
\citet{HasegawaHerbst1993} to grains with $a\in$~[0.03, 0.25]~$\mu$m, 
corresponding to a temperature range of about [200, 35]~K (hence the 
final desorption rate has a weak dependence on $T_{\rm e}$ for a 
given species, consistent with \ct{HasegawaHerbst1993}). 

For larger grains (>0.25~$\mu$m), spot heating by cosmic-rays will start 
to dominate over whole grain heating \citep{Leger+1985, Shen+2004}. 
For smaller grains with much higher $T_{\rm e}$, most CO and other 
volatile species on the grain surface are rapidly depleted through 
evaporation with timescale <10$^{-9}$~s, which only removes 
$\lesssim$10\% of the total deposited energy $E_{\rm dep}$. 
However, because of the high temperature, sublimation of 
H$_2$O becomes very efficient \citep[][each H$_2$O molecule takes 
away five times the energy of each CO molecule]{Tielens2005}. 
The timescale of water sublimation for grains smaller than 0.01~$\mu$m
can be estimated as $\sim$10$^{-8}$~s given $T_{\rm e}$$\sim$600~K. 
For simplicity, these additional processes are not considered in this 
study (detailed treatment of water sublimation on small grains may 
potentially affect the abundances of species with large binding energies). 

Based on these assumption, the duty cycle can be obtained by scaling 
both the timescale of evaporative cooling and the time interval between 
successive heating. The resulting duty cycle $f(T_{\rm e}(a))$ 
can vary by more than 10 orders of magnitude 
(10$^{-26}$--10$^{-12}$~s$^{-1}$) along the standard MRN 
\citep[Mathis-Rumpl-Nordsieck;][]{Mathis+1977} size distribution; 
smaller grains have smaller duty cycles per grain because they are struck 
less frequently by cosmic rays and they spend shorter times at the elevated 
temperature \citep[see also][]{Acharyya+2011}, while larger grains 
have larger cross-section for incoming cosmic rays and they cool down 
slowly since their peak temperatures are closer to the ambient gas 
temperature. 

The final value of <$k_{\rm crd}(i)$> for species $i$ is obtained 
by averaging over the grain size distribution (weighted by grain 
surface area) as:
\begin{equation}
<k_{\rm crd}>={\Sigma_a k_{\rm crd}~\pi a^2 n(a) \over \Sigma_a \pi a^2 n(a)}~,
\end{equation}
which is equivalent to treating each surface species as a separate 
species in each size bin,\footnote{The cosmic-ray desorption term 
in the rate equation Eq.~\ref{Eq:rateEQs} for a size distribution is, 
$$\Sigma_a \Big[k_{\rm crd}(a) {n_{i,s} \over \Sigma_a \pi a^2 n(a)} \pi a^2 n(a)\Big]={\rm <}k_{\rm crd}{\rm >}~n_{i,s}$$} 
given enough surface sites for the adsorbed species at each size bin. 
As an example, the desorption rate $k_{\rm crd}({\rm CO},a)$ for CO takes 
value between $\sim$10$^{-17}$--10$^{-13}$~s$^{-1}$ for the standard MRN  
distribution, with larger values of $k_{\rm crd}$ for larger $a$, 
and the weighted desorption rate 
$k_{\rm crd}(i)$$\approx$1.04$\times$10$^{-14}$~s$^{-1}$ (very similar to 
the values derived by \citet{HasegawaHerbst1993} using 0.1~$\mu$m grains).

Cosmic rays also induce secondary UV photons when interacting with H$_2$ 
molecules -- the so called H$_2$ fluorescence in the Lyman and Wemer bands. 
The resulting flux of UV photons can be estimated by: 
\begin{equation}
F_{\rm UV} \approx 1830 \left({\zeta^{\rm H_2} \over 10^{-17}~s^{-1}}\right)~cm^{-2}~s^{-1}~
\end{equation}
\citep{Cecchi-PestelliniAiello1992,Ivlev+2015b}, assuming the typical 
interstellar dust gas and dust properties. The corresponding rate of 
photo-desorption by such a fluorescence can be expressed as, 
\begin{equation}
k_{\rm crd,fl}={F_{\rm UV} \over 4 n_{\rm s}N_{\rm l}}Y_i~,
\end{equation}
where $N_{\rm l}$=2 is the number of ice layers that can be affected by 
an incoming UV photon for photo-desorption, and $Y_i$ is the 
photo-desorption yield per photon. For simplicity, we set 
$Y_i$=10$^{-3}$ for all desorbing species. 

The coupled ODEs (ordinary differential equations) are solved using DVODE 
library\footnote{available at \url{http://www.radford.edu/~thompson/vodef90web/}}
with sparse matrix turned off, due to the small number of species 
yet a large number of reactions for each species.

\section{Result}
\label{Chap.Result}

We evolve the chemical network for 10$^5$ years for the freeze-out 
mechanism to take effect \citep{Flower+2005}, while the ion chemistry 
reaches equilibrium as quickly as a few 10$^1$ years \citep{Caselli+2002b}. 
In the high density regions, we adopt a barotropic equation of state (EOS) 
described in Appendix~\ref{App.A} to mimic the change of temperature at 
high densities including those for protostellar disks. The cosmic-ray 
ionization rate ($\zeta_0^{\rm H_2}=1.3\times10^{-17}$~s$^{-1}$) 
also attenuates exponentially at high densities following the 
relation given in \ct{Nakano+2002} 
(see also, Eq.~1 in \ct{Zhao+2016}), until a lower limit of 
$1.1\times10^{-22}$~s$^{-1}$ (corresponding to the ionization 
by radioactive decay of long-lived $^{40}$K) is reached. 

In this section, we demonstrate that grain size distribution can 
significantly affect cosmic-ray desorption as well as non-ideal MHD 
effects. In particular, 
(1) the abundance differentiation among volatile species is sensitive 
to the maximum grain size ($a_{\rm max}$>0.1~$\mu$m); 
(2) ambipolar diffusivity is sensitive to the amount of VSGs 
\citep[a few to tens of nanometer, see also][]{Zhao+2016}; 
(3) Hall diffusivity is generally sensitive to grain size distribution 
in terms of sign and magnitude, but reaches a maximum level at densities 
below 10$^{13}$~cm$^{-3}$ when $a_{\rm min} \approx 0.04~\mu$m 
\citep[in contrast to][]{Dzyurkevich+2017}.

\subsection{Amplified Differential Depletion of Volatile Species Due to Cosmic-ray Desorption}
\label{S.Nitrogen}

Conventionally, the small difference in binding energies among volatile 
species are considered to have negligible effect on their gas-phase 
abundances. However, with the new formulation of cosmic-ray desorption 
rate which includes the dependence on grain size, the effect of 
binding energies on the abundances of volatiles is amplified by the 
variations in the elevated temperature. As a result, the small difference 
in binding energies can be the origin of the differential depletion among 
volatiles in this paradigm.\footnote{Indeed, more complete chemical 
networks including grain surface reactions are needed to 
fully understand the differential depletion of volatile species 
in dense molecular clouds; we will leave it to future investigations.}

\begin{figure*}
\includegraphics[width=\textwidth]{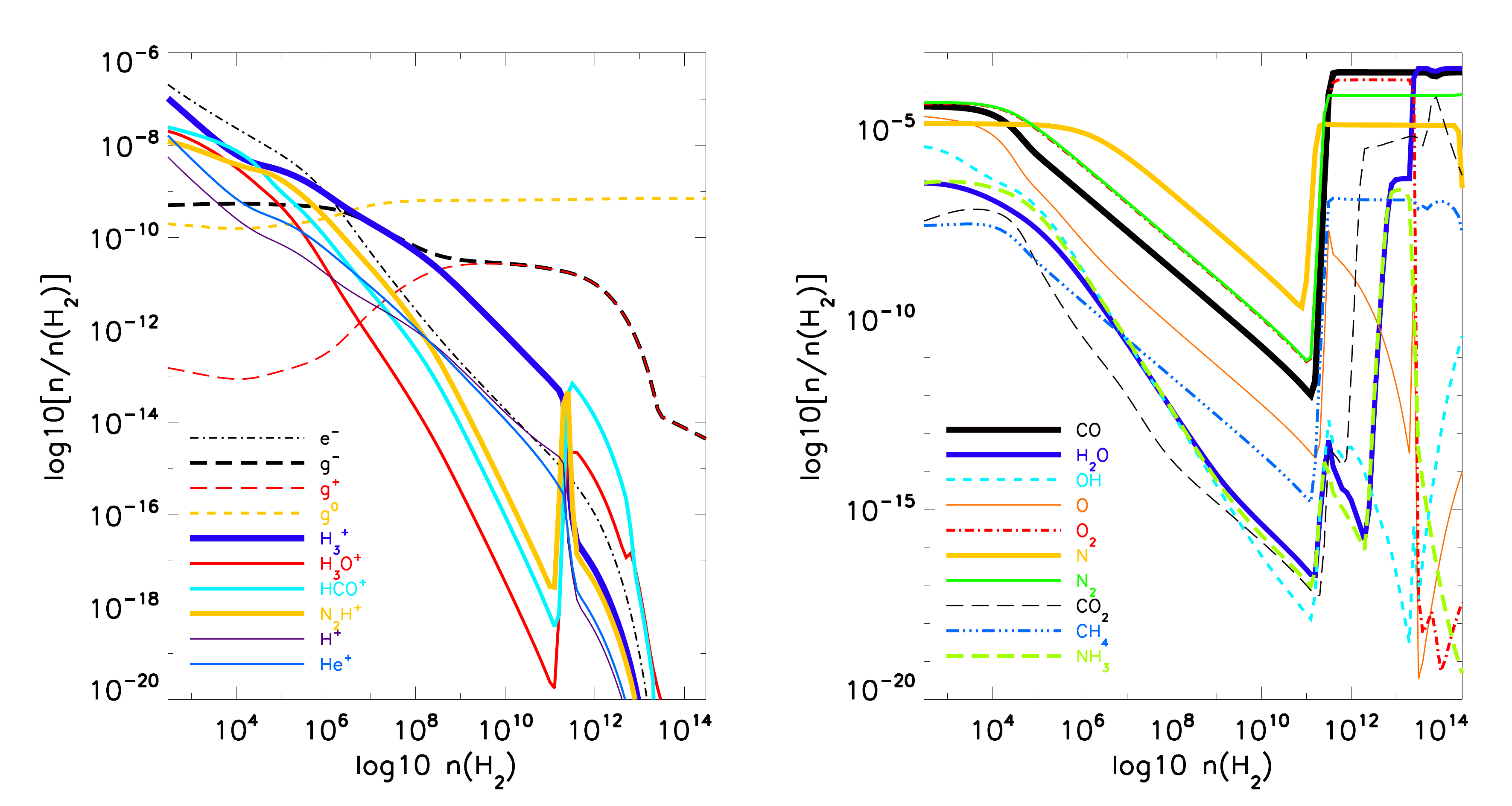}
\caption{Fractional abundances of the main ion and neutral species 
computed with the new formulation of cosmic-ray desorption rate for a 
standard MRN size distribution. In the right panel, N atom is $\sim$2 
orders of magnitude more abundant than CO for number densities between 
10$^5$--10$^{11}$~cm$^{-3}$.}
\label{Fig:fld+crd_MRN}
\end{figure*}

\begin{figure*}
\includegraphics[width=\textwidth]{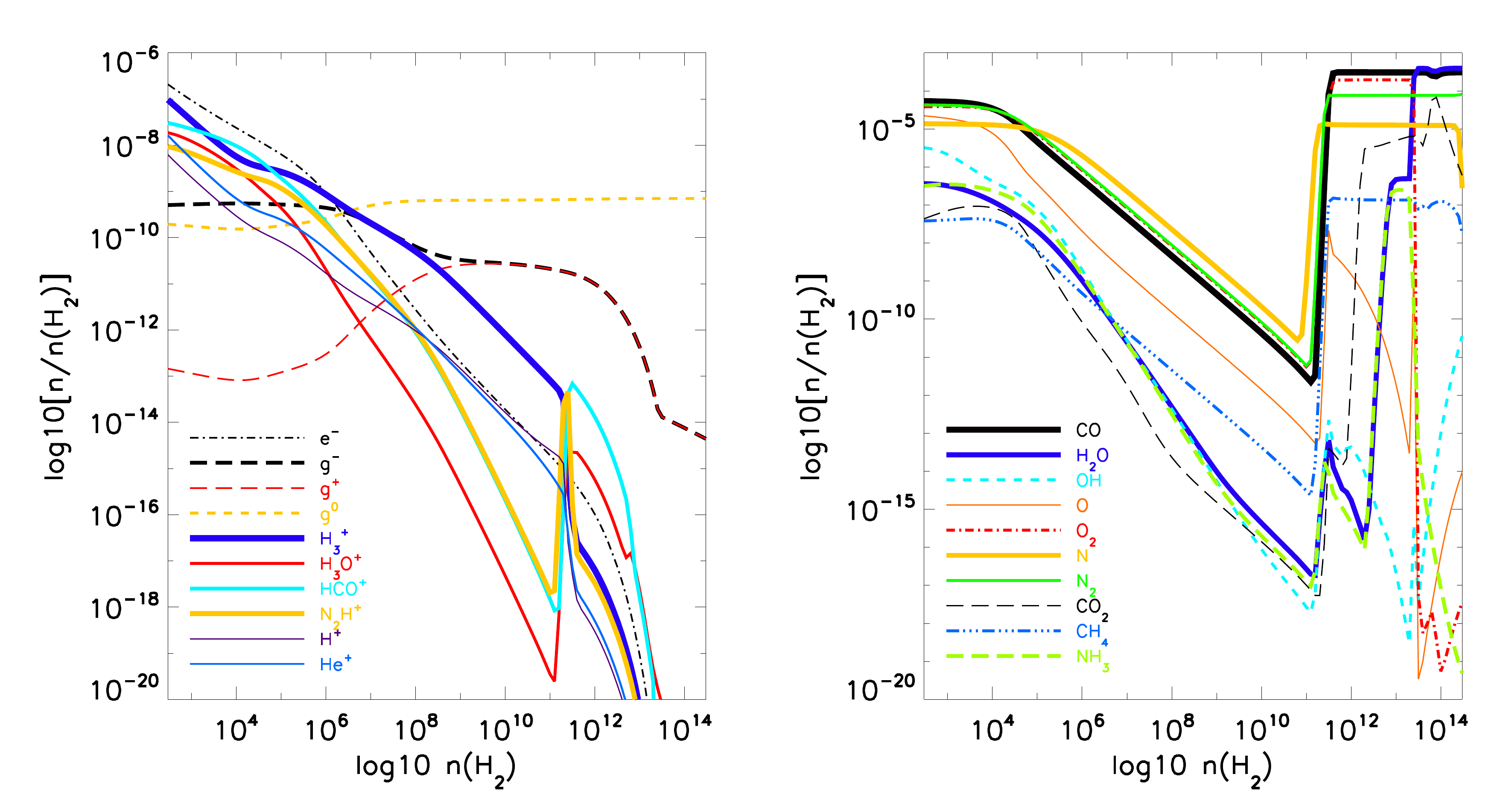}
\caption{Fractional abundances of the main ion and neutral species 
computed with the conventional $f$(70~K) pre-factor. Other parameters 
remain the same as in Fig.~\ref{Fig:fld+crd_MRN}. The abundances of 
volatile species in the right panel are nearly indistinguishable 
from each other.}
\label{Fig:fld+crd_MRN_f70}
\end{figure*}

Fig.~\ref{Fig:fld+crd_MRN} shows the abundances of the main 
ion and neutral species relative to H$_2$ using the new formulation 
of cosmic-ray desorption rate (Eq.~\ref{Eq:k_crd}). With the standard MRN 
size distribution, the gas-phase atomic N is two orders 
of magnitude more abundance than gas-phase CO (in the density range 
between a few $10^5$ and a few $10^{10}$~cm$^{-3}$), while the 
abundances of N$_2$ and O$_2$ are higher than CO by a factor of few 
(consistent with observational constraints by \ct{Maret+2006}; 
see also \ct{vanDishoeckBlake1998}). The excess of N$_2$ over CO 
in the gas phase also leads to an excess of N$_2$H$^+$ over HCO$^+$ 
by a factor of a few in the density range of dense cores. 
In contrast, when the conventional pre-factor for the cosmic-ray 
desorption rate $f(70~{\rm K})=3.16\times10^{-19}$ (derived with 0.1~$\mu$m 
grains) is used instead, the difference in abundances among volatile 
species becomes almost negligible (Fig.~\ref{Fig:fld+crd_MRN_f70}), 
which recovers the claims of existing studies \citep[e.g.,][]{Flower+2005}. 

\begin{figure*}
\includegraphics[width=\textwidth]{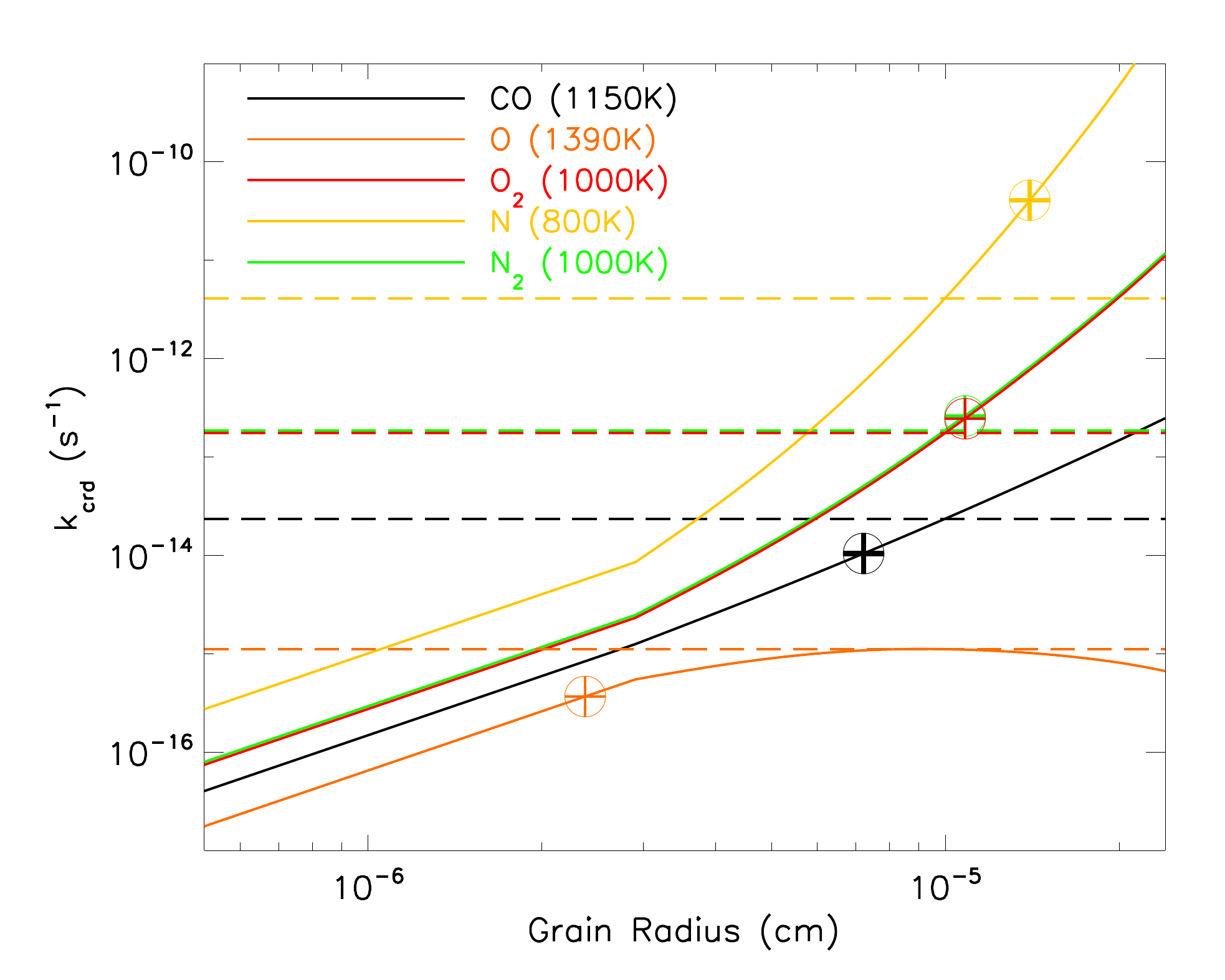}
\caption{Comparison of cosmic-ray desorption rates among volatile species. 
Solid lines represent the desorption rate per grain $k_{\rm crd}(i,a)$ for 
a given species $i$ and grain radius $a$. 
The circled crosses mark the averaged desorption rate for each species. 
Long dashed lines are the desorption rate computed using the conventional 
$f$(70~K) pre-factor. The value of binding energies are taken from 
\citet{Sipila+2015}.}
\label{Fig:compare_crd}
\end{figure*}

The differential depletion of volatile species is clearly amplified 
by adopting the new cosmic-ray desorption rate that varies with grain sizes. 
Fig.~\ref{Fig:compare_crd} compares for the main volatile species 
the cosmic-ray desorption rates $k_{\rm crd}(i,a)$ 
(computed via Eq.~\ref{Eq:k_crd}--\ref{Eq:E_dep}) and the averaged value 
<$k_{\rm crd}(i)$>, along with the conventional rate for 0.1~$\mu$m grain 
from \citet{HasegawaHerbst1993}. Apparently, the averaged desorption rates 
<$k_{\rm crd}(i)$> (circled crosses in Fig.~\ref{Fig:compare_crd}) 
of different volatile species are more separated from each other than 
the conventional desorption rates computed using 0.1~$\mu$m grain (dashed 
horizontal lines). The enhanced differential depletion is caused by 
the large variation of $k_{\rm crd}(i,a)$ (among different species) 
in the grain size range larger than $\sim$0.1~$\mu$m; these grains are 
only heated by cosmic rays to lower elevated temperatures $T_{\rm e}(a)$ 
(e.g. $\sim$40~K) and are more sensitive to the difference in binding 
energies among volatile species. For example, $k_{\rm crd}(i,0.2~\mu {\rm m})$ 
of volatile species with small binding energies (e.g., N and N$_2$) reaches 
$\sim$10$^{-11}$--10$^{-10}$~s$^{-1}$, compared to 10$^{-15}$~s$^{-1}$ for 
less volatile species (e.g., O). Therefore, the averaged <$k_{\rm crd}(i)$> 
(weighted by grain surface area) lean towards larger values for more 
volatile species (e.g., N and N$_2$) but smaller values for 
less volatile species (e.g., O, H$_2$O). We have also tested models 
with longer evolution time 10$^6$~yrs and 10$^7$~yrs, and the effect 
of differential depletion is almost identical. 

Note that on the freeze-out timescale ($\sim$10$^5$~yrs), 
the majority of desorption of volatile species 
($E_{\rm des}$$\sim$1000~K such as CO, N$_2$, O$_2$, C, N, and O) 
owes to the impulsive heating by cosmic-rays. 
In comparison, the cosmic-ray induced UV photons contribute most to 
the desorption of species such as OH and H$_2$O that have large 
binding energies $E_{\rm des}$. Indeed, $k_{\rm crd}(i,a)$ is a 
decreasing function of $a$ at larger grain sizes, 
because the evaporative cooling timescale is derived by assuming CO as 
the primary volatile species to cool the grains 
\citep[][]{HasegawaHerbst1993}, and within the duty cycle, the 
non-volatile species are difficult to evaporate into gas-phase when 
the elevated temperature is low. A more strict treatment for 
the evaporative cooling process including the contributions from 
both CO and H$_2$O ice may provide more accurate results for these 
non-volatile species. 
Again, we are aware of the disagreements over binding energies 
among existing literature, and improved measurements of 
binding energies \citep[e.g.][]{Cazaux+2017,Shimonishi+2018} are 
needed to refine the cosmic-ray desorption model presented here.

\subsection{Freeze-Out and Relative Abundance of Ions}
\label{S.Freeze}

\begin{figure*}
  \centering
  \subfloat[Freeze-out and desorption: Off]{
    \includegraphics[width=\columnwidth]{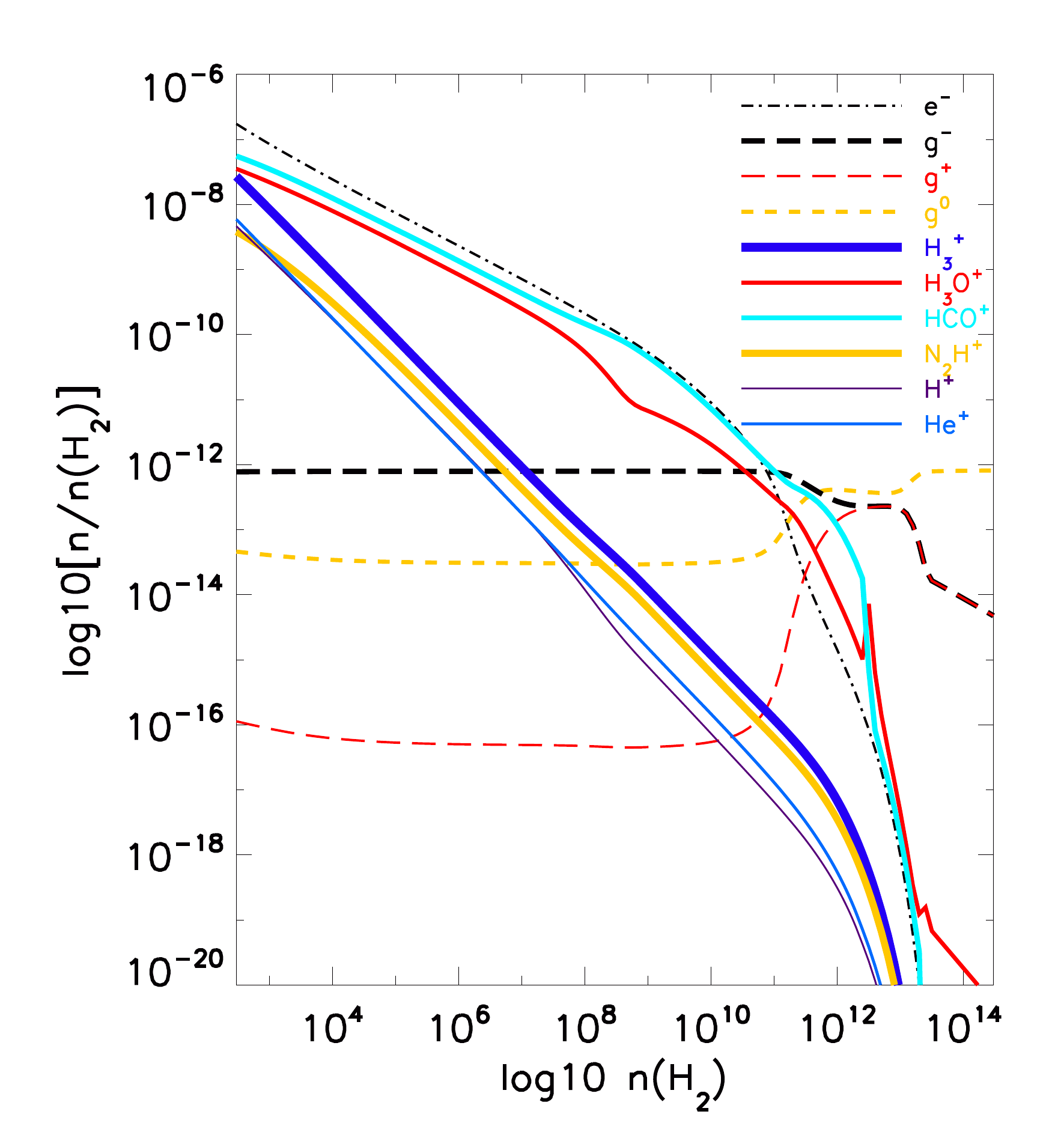}
    \label{Fig:Ion-trMRN-noFD}
  }
  \hfill
  \subfloat[Freeze-out and desorption: On]{
    \includegraphics[width=\columnwidth]{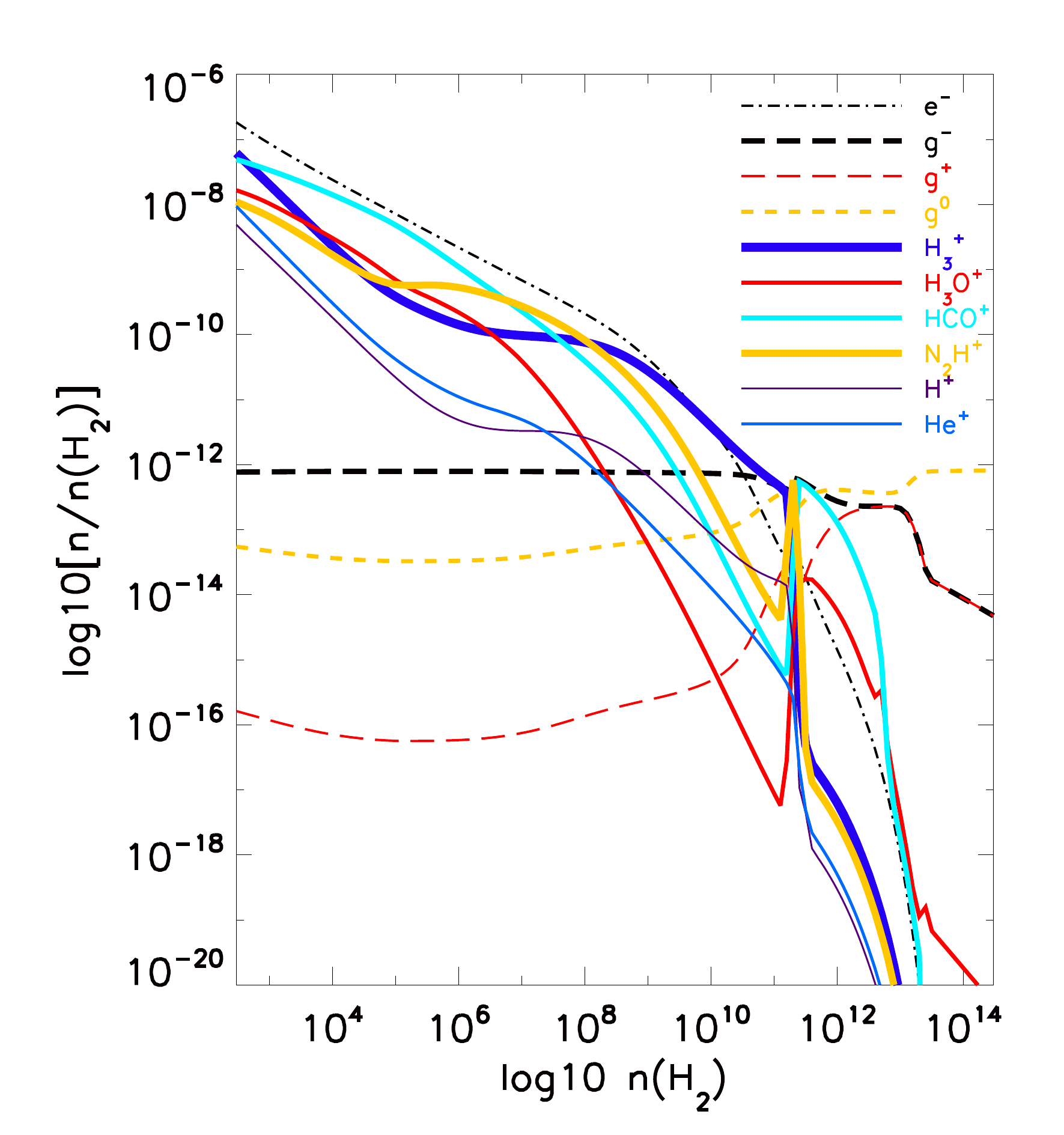}
    \label{Fig:Ion-trMRN-FD}
  }
  \caption{Fractional abundances of the main ion species computed without 
    freeze-out/desorption (left panel) and with freeze-out/desorption 
    (right panel) for the truncated MRN size distribution with 
    $a_{\rm min}$=0.1~$\mu$m.}
\end{figure*}

As shown in \S~\ref{S.Nitrogen} above, the chemical abundances are 
strongly affected by the freeze-out and desorption processes. 
In particular, the dominant ion species in dense cores is no 
longer the commonly assumed HCO$^+$ \citep[Fig.~\ref{Fig:Ion-trMRN-noFD};
see also][]{UmebayashiNakano1990,Zhao+2016,Marchand+2016} 
but H$_3^+$ instead for number densities below 10$^{11}$~cm$^{-3}$ 
for the standard MRN size distribution (see Fig.~\ref{Fig:fld+crd_MRN} 
above), consistent with the result of \citet{Flower+2005}.

However, the grain size distribution can also affect the relative 
abundances among ion species via the change of surface area for 
freeze-out. As shown in Fig.~\ref{Fig:Ion-trMRN-FD}, when using 
a truncated MRN distribution with $a_{\rm min}$=0.1~$\mu$m, 
H$_3^+$ only starts to dominate the ion abundances for densities between 
10$^8$--10$^{11}$~cm$^{-3}$. In the low density regime, 
HCO$^+$ is still the most abundant ion species up to 
$\sim$8$\times$10$^6$~cm$^{-3}$, when N$_2$H$^+$ overtakes HCO$^+$. 
In comparison, with the standard MRN distribution 
(Fig.~\ref{Fig:fld+crd_MRN}), N$_2$H$^+$ starts to dominate over 
HCO$^+$ at relatively low densities $\sim$5$\times$10$^4$~cm$^{-3}$, 
but they both are much less abundant than H$_3^+$ 
(10$^5$--10$^{11}$~cm$^{-3}$~cm$^{-3}$). 
It is likely that neither size distributions can well characterize the 
grain properties in dense cores over the long time evolution 
\citep[e.g.,][]{Hirashita2012}. We also show an intermediate case 
with $a_{\rm min}$=0.04~$\mu$m in Appendix~\ref{App.C}, 
the dominant ion species switches from HCO$^+$ to N$_2$H$^+$ at 
number density around 8$\times$10$^5$~cm$^{-3}$, and H$_3^+$ dominates 
over both between 5$\times$10$^6$--10$^{11}$~cm$^{-3}$. The change of 
the dominant ion species is regulated by the level of depletion of 
volatile species (e.g., CO and N$_2$), which is sensitive to the 
grain surface area for freeze-out.

Observationally, the relative abundances among the main ion species 
may provide an alternative way of determining the average grain 
size <$a$> in dense cores. For instance, if the abundance of H$_3^+$ 
could be inferred either directly or via comparison of H$_2$D$^+$ 
and other molecular ion observations with chemical models, 
one can obtain an 
estimate of the grain size by comparing the abundance of H$_3^+$ with 
that of N$_2$H$^+$ (or HCO$^+$) at given densities (slightly larger 
<$a$> makes it harder for H$_3^+$ to dominate over the other two ions). 
Furthermore, with a proper limit on the maximum grain size constrained 
by existing observations (if any), the minimum grain size can be inferred 
(assuming a certain power law index for the size distribution), 
and vice versa.

\subsection{Effect of Chemistry on Non-ideal MHD Diffusivities}
\label{S.GrainSize}

With the ionization fraction from the chemical network, it is 
straightforward to obtain the three non-ideal MHD diffusivities: 
$\eta_{\rm AD}$, $\eta_{\rm Ohm}$, $\eta_{\rm Hall}$ 
\citep[Appendix~\ref{App.B}; see also,][]{WardleNg1999,Zhao+2016}. 
Despite the pronounced effect of freeze-out and desorption processes 
on the abundance of ions, they only have limited influence on the 
non-ideal MHD effects that are instead modulated by the grain size 
distribution \citep{WardleNg1999,Padovani+2014,Zhao+2016,Dzyurkevich+2017}. 
However, disagreement still exists among literatures in terms of the 
effect of grain size on magnetic diffusivities. We now utilize our 
chemical network to clarify the disagreement, with a particular 
emphasis on Hall diffusivity, which is not covered in details 
in \citet{Zhao+2016}. 

\subsubsection{Limited Influence of Freeze-out and Desorption on Magnetic Diffusivities}
\label{S.ChemOnMHD}

\begin{figure*}
\includegraphics[width=1.03\textwidth]{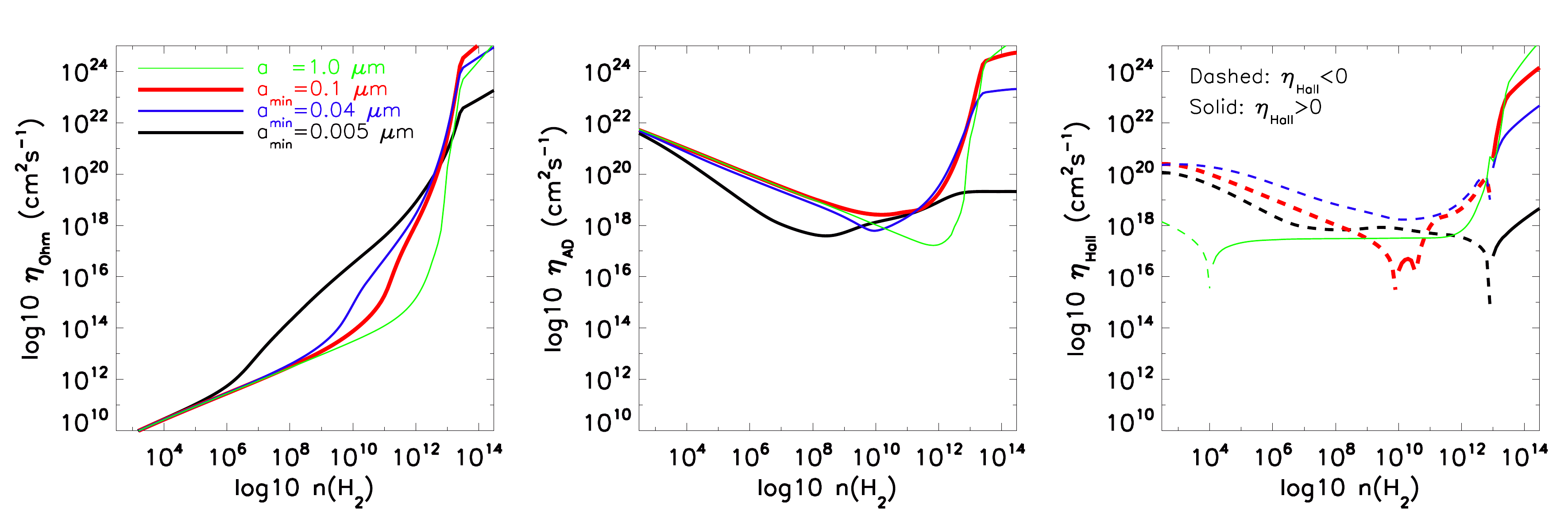}
\includegraphics[width=1.03\textwidth]{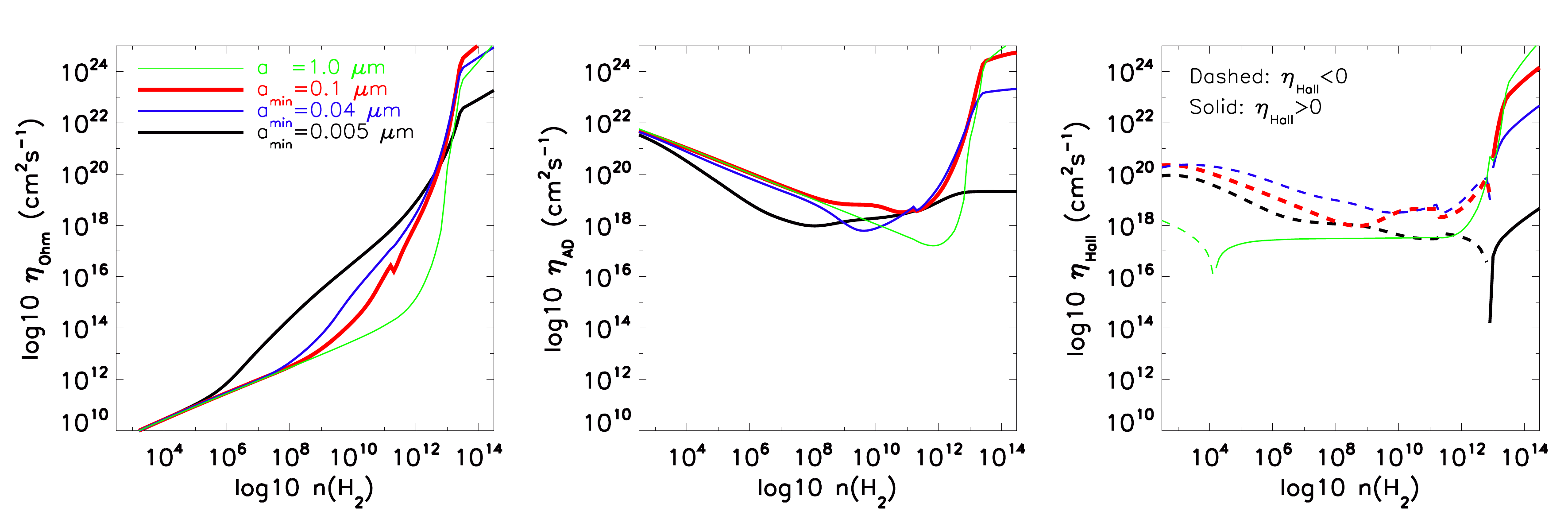}
\caption{Non-ideal MHD diffusivities computed without (top three panels) 
and with (bottom three panels) freeze-out/desorption for different size 
distributions with $a_{\rm min}=0.005~\mu$m 
(MRN: black), $a_{\rm min}=0.04~\mu$m (blue), $a_{\rm min}=0.1~\mu$m (red), 
$a=1.0~\mu$m (green). Ambipolar diffusion is at a maximum level over 
the density range when $a_{\rm min}=0.1~\mu$m, while Hall effect is 
more efficient for disk formation when $a_{\rm min}=0.04~\mu$m.}
\label{Fig:compare_etas}
\end{figure*}

We again vary the minimum grain size $a_{\rm min}$ of the size distribution, 
but keep the power index of -3.5 and $a_{\rm max}=0.25~\mu$m. 
Similar to \citet{Li+2011} and \citet{Zhao+2016}, we adopt the simple 
relation between magnetic field and density \citep{Nakano+2002}, 
\begin{equation}
\label{Eq:Brelation}
|\bmath{B}|=0.143~\left[{n({\rm H}_2) \over {\rm cm}^{-3}}\right]^{0.5}~\mu{\rm G}~,
\end{equation}
to estimate the magnetic field strength. 
The resulting magnetic diffusivities for different grain sizes are 
shown in Fig.~\ref{Fig:compare_etas}, with freeze-out/desorption 
turned on or off. 
Overall, the magnetic diffusivities are very similar between the two 
cases, indicating that changes in ion abundances\footnote{The change 
in Hall parameter (defined in Appendix~\ref{App.B}) by switching the 
dominant ion species from HCO$^+$ to H$_3+$ has even less effect 
($\lesssim$~5\%).} due to freeze-out and/or desorption only have 
limited effect on the magnetic diffusivities. 

\begin{figure*}
\includegraphics[width=\textwidth]{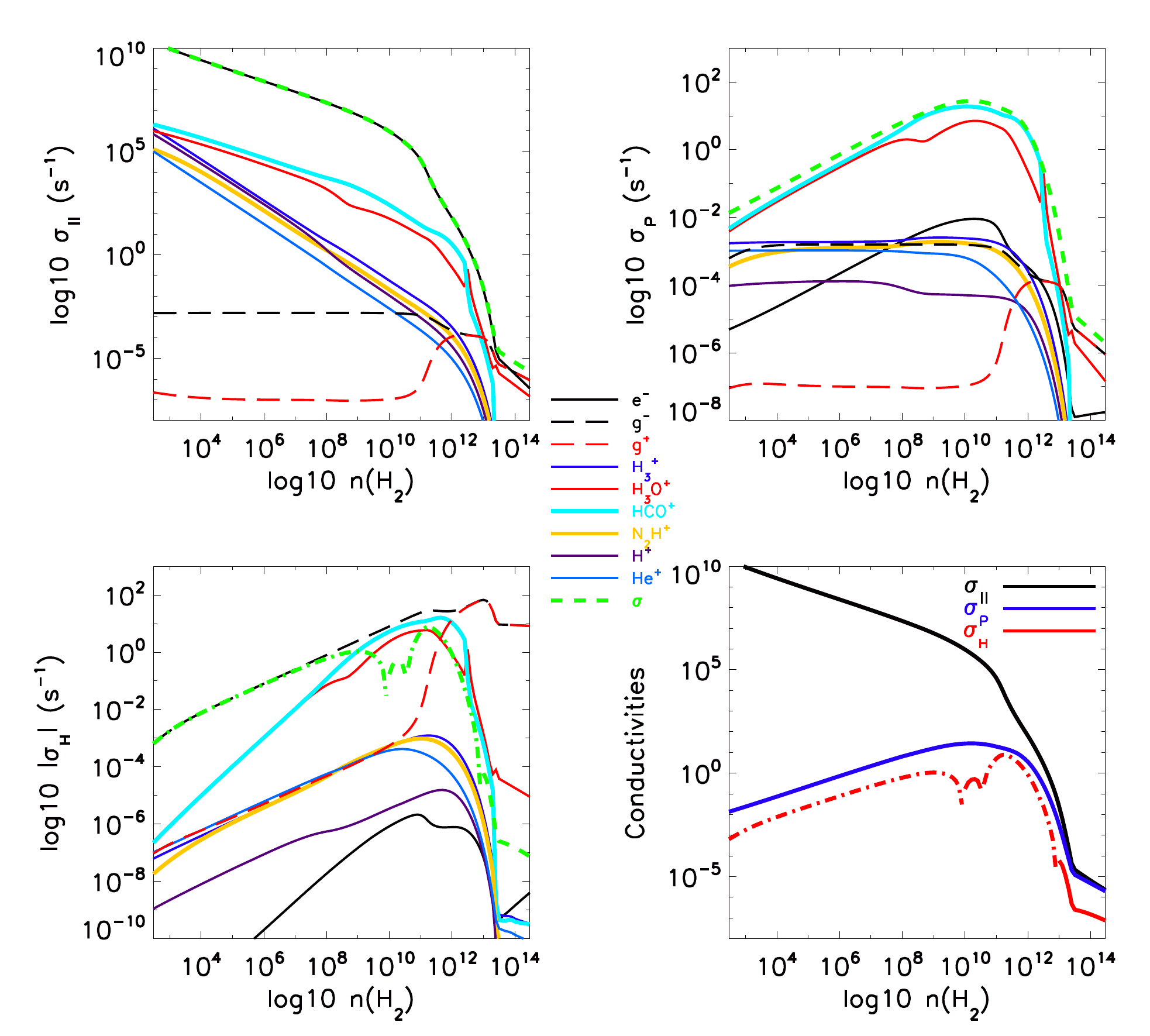}
\caption{Components of conductivity tensor and the contributions from 
major charged species, for the truncated MRN size distribution with 
$a_{\rm min}=0.1~\mu$m and $a_{\rm max}=0.25~\mu$m, with freeze-out 
and desorption turned off.}
\label{Fig:UNcond_1k}
\end{figure*}

\begin{figure*}
\includegraphics[width=\textwidth]{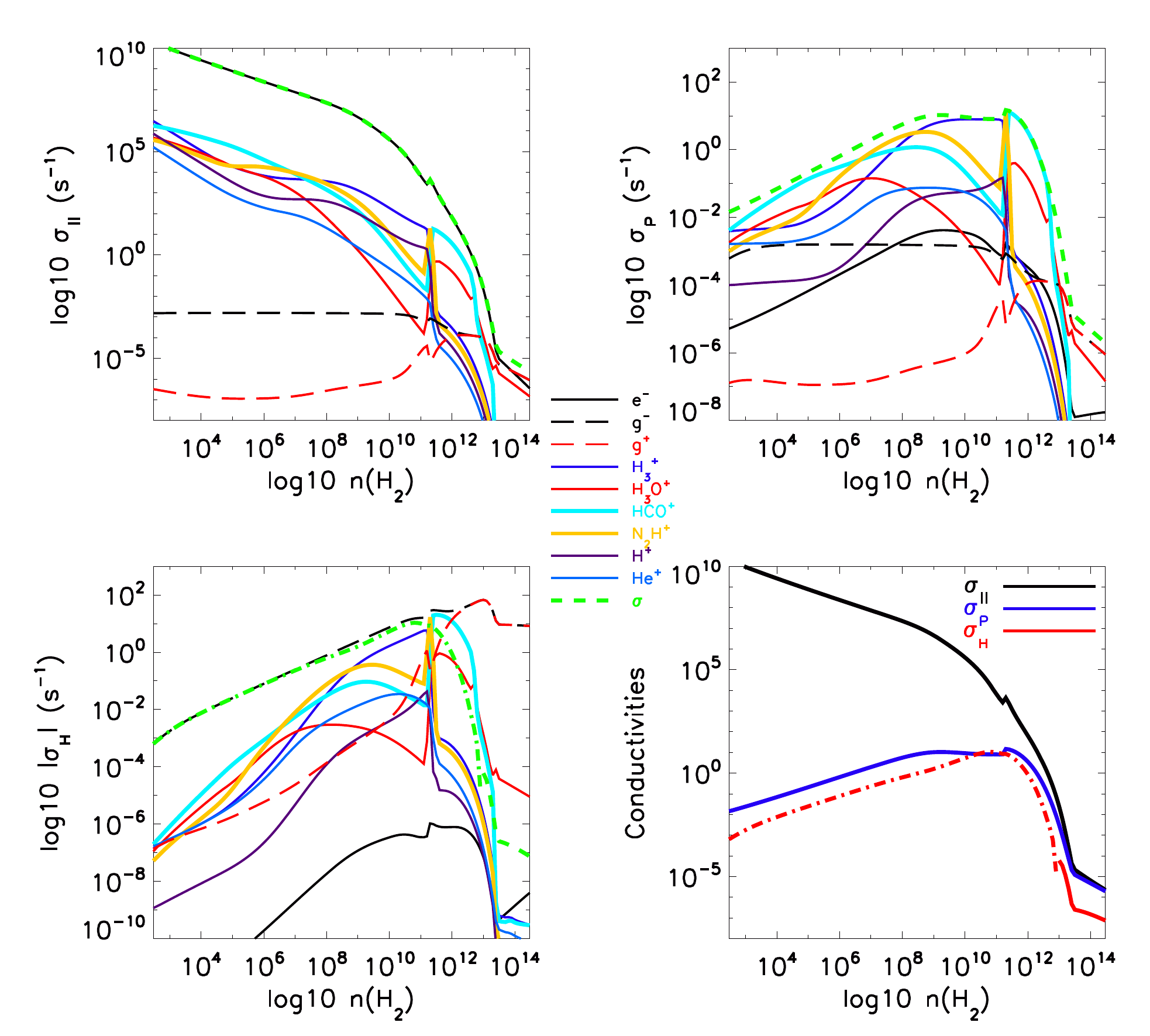}
\caption{Same as Fig.~\ref{Fig:UNcond_1k}, but with freeze-out and desorption
turned on.}
\label{Fig:hcond_1k}
\end{figure*}

The most obvious difference between the two cases is the Hall diffusivity 
in the density range $\sim$10$^9$--10$^{11}$~cm$^{-3}$ when using 
$a_{\rm min}=0.1~\mu$m. In the case without freeze-out and desorption, 
$\eta_{\rm Hall}$ briefly switches to positive sign in the vicinity 
of 10$^{10}$~cm$^{-3}$, causing a wider density range 
($\sim$10$^9$--10$^{11}$~cm$^{-3}$) to have lowered values of 
$\eta_{\rm Hall}$. In contrast, the case with freeze-out 
and desorption (bottom panels of Fig.~\ref{Fig:compare_etas}) shows a 
more smooth negative value of $\eta_{\rm Hall}$ in the similar 
density range. The reason for such a difference is evident when comparing 
Fig.~\ref{Fig:UNcond_1k} and Fig.~\ref{Fig:hcond_1k} where we plot the 
contributions of the main charged species to the three components 
of fluid conductivity (parallel $\sigma_{\parallel}$, 
Pedersen $\sigma_{\rm P}$, and Hall $\sigma_{\rm H}$ defined in 
Appendix~\ref{App.B}). The principal term that determines Hall 
diffusivity is the Hall conductivity $\sigma_{\rm H}$, which carries 
a sign for each charged species (Eq.~\ref{Eq:conduct3}). In the 
case without freeze-out and desorption, $\sigma_{\rm H}$ 
becomes positive near 10$^{10}$~cm$^{-3}$ because the total 
contribution of $\sigma_{\rm H}$ from ions (especially HCO$^+$ 
and H$_3$O$^+$) surpasses $\sigma_{\rm H}$(g$^-$). However, when 
freeze-out and desorption are turned on, $\sigma_{\rm H}$ is always 
dominated by g$^-$ in low density regimes ($\lesssim$10$^{11}$~cm$^{-3}$). 
In fact, when we adopt $a_{\rm min}\approx0.15~\mu$m for the 
freeze-out/desorption case, $\eta_{\rm Hall}$ presents similar 
sign change between 10$^{9}$--10$^{10}$~cm$^{-3}$. 
Therefore, it is a general behavior of Hall diffusivity that rather 
depends on the choice of grain size than on freeze-out/desorption. 
As grain size increases, the abundance of g$^-$ reduces, so 
that positive Hall diffusivity extends to wider and wider regions 
in the lower density regime ($\lesssim$10$^{11}$~cm$^{-3}$). 

\subsubsection{Optimal Grain Size for Hall Diffusivity}
\label{S.OptimalHall}

\begin{figure*}
\includegraphics[width=\textwidth]{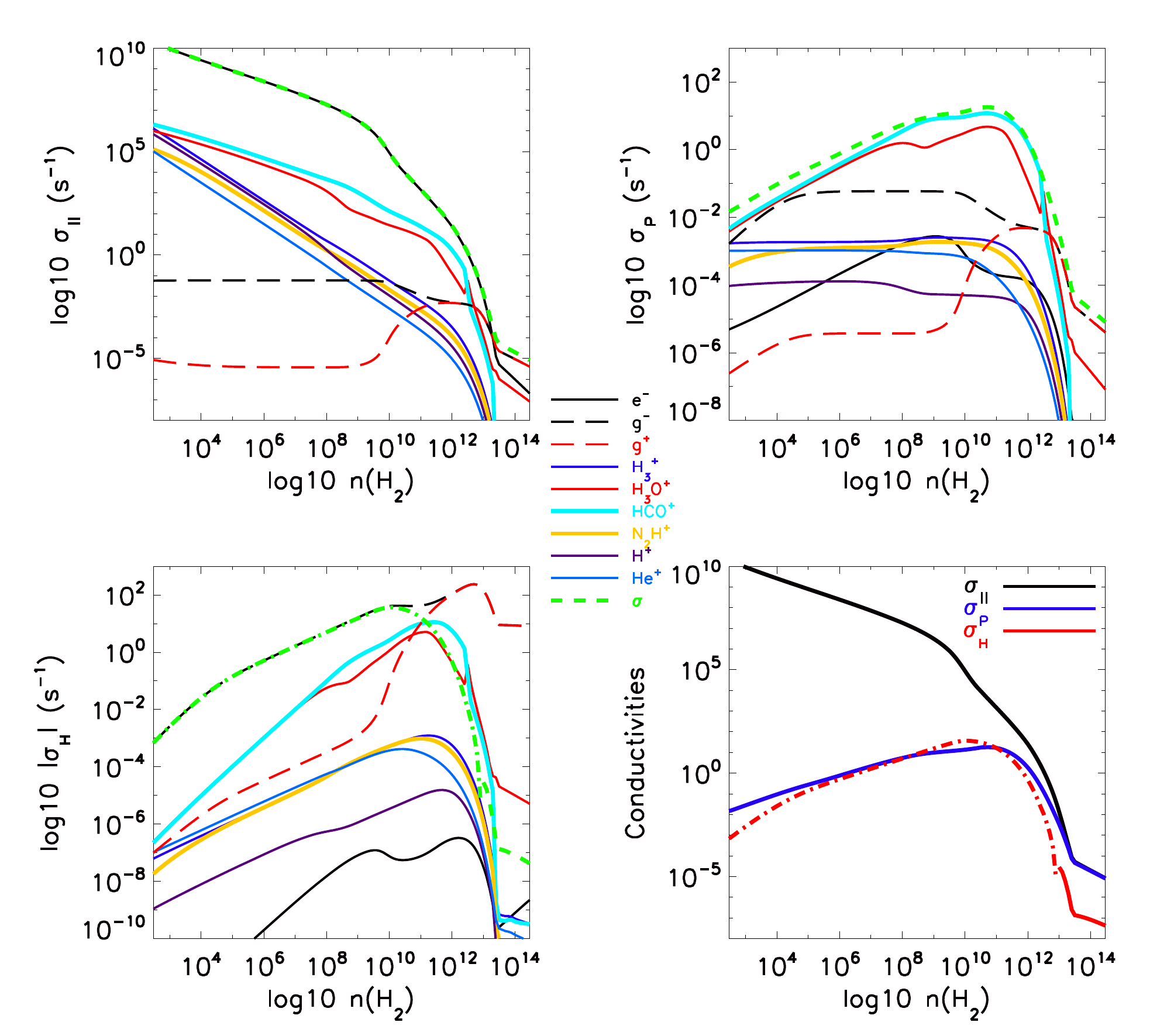}
\caption{Components of conductivity and the contributions from major 
charged species, for the truncated MRN size distribution with 
$a_{\rm min}=0.04~\mu$m and $a_{\rm max}=0.25~\mu$m. This size distribution 
produces the strongest Hall effect (at densities $\lesssim10^{11}$~cm$^{-3}$). 
Freeze-out and desorption are turned off.}
\label{Fig:UNcond_400}
\end{figure*}

In fact, there exists an optimal grain size $a_{\rm min}\approx0.04~\mu$m 
for which Hall diffusivity $\eta_{\rm Hall}$ reaches a maximum level 
at number densities below $\lesssim$10$^{11}$~cm$^{-3}$. 
According to Eq.~\ref{Eq:MHDcoef3} in Appendix~\ref{App.B}, large 
$\eta_{\rm Hall}$ requires both Hall conductivity $\sigma_{\rm H}$ 
and Pedersen conductivity $\sigma_{\rm P}$ to be small, and at the 
same time $\sigma_{\rm H}\geqslant\sigma_{\rm P}$. As shown in 
Fig.~\ref{Fig:UNcond_400} for $a_{\rm min}\approx0.04~\mu$m, the 
two conductivity components $\sigma_{\rm H}$ and $\sigma_{\rm P}$ 
are roughly equal to each other ($\lesssim$10$^{11}$~cm$^{-3}$). 
The Hall diffusivity $\eta_{\rm Hall}$ reaches a maximum here because 
(1) for larger $a_{\rm min}$ ($\gtrsim$0.04~$\mu$m), $\sigma_{\rm H}$ 
starts to decrease below $\sigma_{\rm P}$, and (2) for smaller 
$a_{\rm min}$ ($\lesssim$0.04~$\mu$m), $\sigma_{\rm H}$ increases first 
and followed by the increase of $\sigma_{\rm P}$ ($\lesssim$0.01~$\mu$m). 
$\eta_{\rm Hall}$ tends to be reduced either way. 
The key is that Hall conductivity $\sigma_{\rm H}$ generally 
increases with decreasing abundance of g$^{-}$, and Pederson 
conductivity $\sigma_{\rm P}$ starts to be significantly 
dominated by VSGs when $a_{\rm min}\lesssim0.02~\mu$m. 
Note that the trend presented here is essentially the same as that 
shown in Appendix A of \citet{Zhao+2016}, except that the main focus 
was previously ambipolar diffusivity.

At high densities ($\gtrsim$10$^{13}$~cm$^{-3}$), the Hall diffusivity 
is mostly positive and increases monotonically with increasing grain size. 
More specifically, the overall magnitude of Hall diffusivity at densities 
$\gtrsim$10$^{11}$~cm$^{-3}$ (protostellar disk density) is quite low 
when $a_{\rm min}\lesssim0.01\mu$m; and it increases by a few orders of 
magnitude when $a_{\rm min}\gtrsim0.02\mu$m. It is consistent with 
the behaviors of Pederson and Hall conductivities in this grain size 
range. 

\subsubsection{Comparison with Other Work}
\label{S.compareHall}

\begin{figure}
\includegraphics[width=\columnwidth]{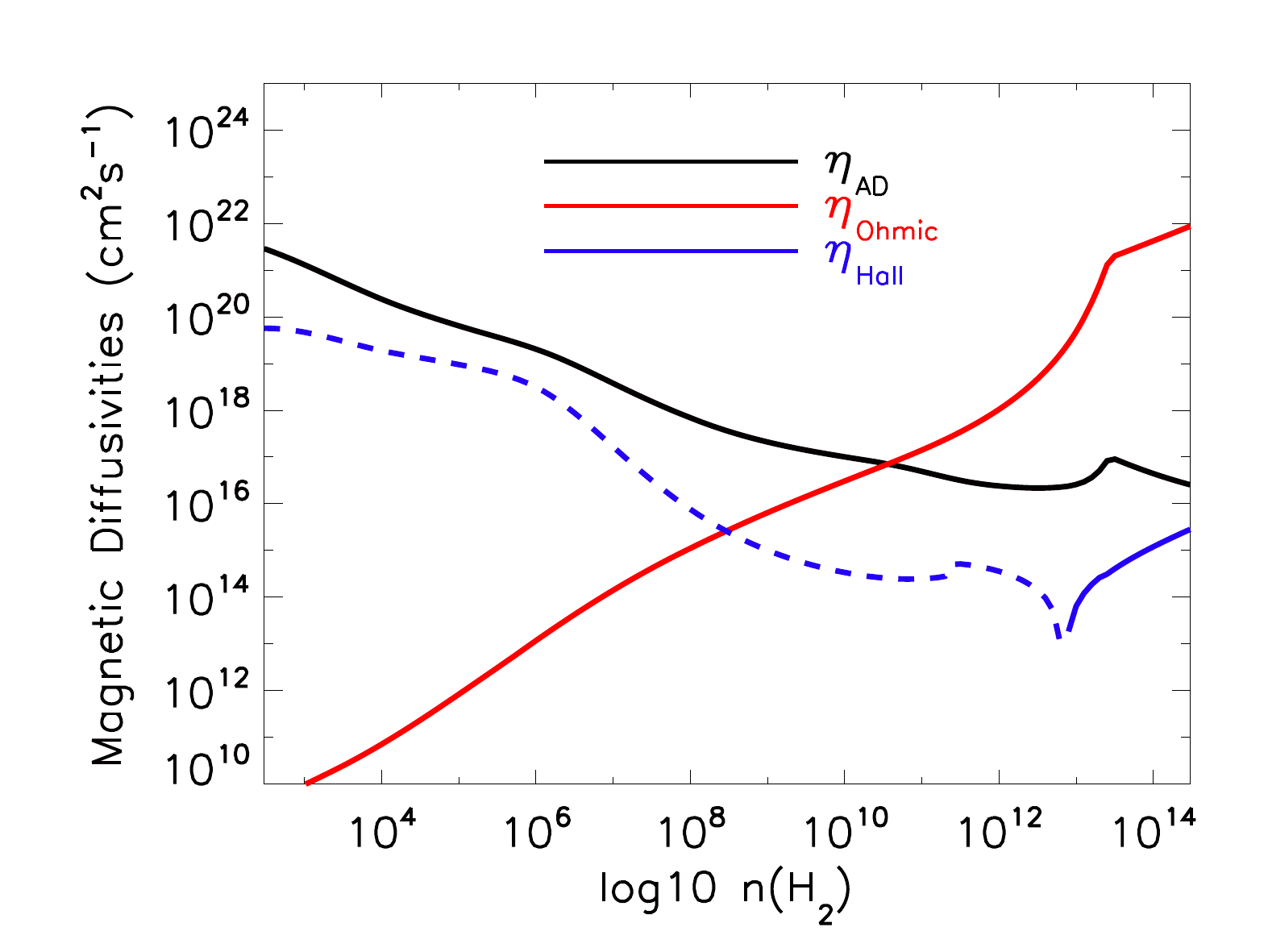}
\caption{Magnetic diffusivities computed with freeze-out/desorption 
for $a_{\rm min}=10~\AA$ (mimic a population of PAHs).} 
\label{Fig:hcno_etas10A}
\end{figure}

Comparing our result of Hall diffusivity with that of 
\citet{Dzyurkevich+2017}, we agree on the fact that larger grain 
sizes (>0.1~$\mu$m) greatly reduces the Hall diffusivity due to 
the reduction of the abundance of $g^-$ and hence $\sigma_{\rm H}$(g$^-$). 
However, we do notice a few differences:
\begin{description}
\item(1) The existence of a maximum level of Hall diffusivity (for number 
densities $\lesssim10^{11}$~cm$^{-3}$) when $a_{\rm min}\approx0.04~\mu$m 
(or <$a$>$\approx$0.063~$\mu$m) is not discussed in \citet{Dzyurkevich+2017}.
\item(2) The sign of Hall diffusivity becoming negative (for number densities 
$\lesssim10^{11}$~cm$^{-3}$) applies to all cases with 
$a_{\rm min}\lesssim0.1~\mu$m (or <$a$>$\approx$0.14~$\mu$m). 
In constrast, \citet{Dzyurkevich+2017} show that the behavior 
of Hall diffusivity $\eta_{\rm Hall}$ (as well as $\eta_{\rm Ohmic}$ and 
$\eta_{\rm AD}$) in the cases with <$a$>=0.05~$\mu$m and <$a$>=0.1~$\mu$m 
are almost identical to each other (their Fig.~2), which is different 
from what we found using our network: 
$\eta_{\rm Hall}$ in their two cases differs by more than 1 order of 
magnitude; the sign change 
only occurs in the case with <$a$>=0.1~$\mu$m but not for <$a$>=0.05~$\mu$m. 
\item(3) The statement in \citet{Dzyurkevich+2017} that Hall effect 
always dominates ambipolar diffusion for grain sizes below\footnote{We 
obtain a value of $a_{\rm min}$$\sim$0.03~$\mu$m (or <$a$>=0.048~$\mu$m) 
for the similar effect, but the Hall effect only dominate over AD in a 
limited density range near 10$^8$~cm$^{-3}$.} 0.02~$\mu$m is no longer 
true when $a_{\rm min}<0.004~\mu$m. 
Recall that (\S~\ref{S.OptimalHall}; see also \ct{Zhao+2016}) 
$\sigma_{\rm H}$ starts to increase when $a_{\rm min}$ drops below 
$0.04~\mu$m, yet $\sigma_{\rm P}$ increases more rapidly when 
$a_{\rm min}$<0.02~$\mu$m and becomes larger than $\sigma_{\rm H}$ 
when $a_{\rm min}<0.004~\mu$m. Therefore, AD regains the dominance 
over Hall effect as tiny grains with size comparable to PAHs are 
included, but neither diffusivities are large enough for efficient 
magnetic decoupling. For example, Fig.~\ref{Fig:hcno_etas10A} shows 
the magnetic diffusivities for a size distribution with 
$a_{\rm min}=10~\AA$, where both $\eta_{\rm AD}$ and $\eta_{\rm Hall}$ 
are orders of magnitude smaller than the models with larger 
$a_{\rm min}$ (Fig.\ref{Fig:compare_etas}). 
\end{description}

Besides the Hall diffusivity, the result of Ohmic and ambipolar 
diffusivity (first two panels of Fig.~\ref{Fig:compare_etas}) is 
essentially the same as in \citet{Zhao+2016} computed using a more 
simplified network (see their \S~4.2, \S~4.4, and Appendix A). 
The main result remains the same: removal of VSGs (a few to tens of 
nanometer) significantly enhances the ambipolar diffusivity. 
Note that we incorporated a barotropic EOS in this work, hence 
the high temperature in the high density regime 
($\gtrsim10^{11}$~cm$^{-3}$) causes the magnetic diffusivities to be 
higher than the values shown in \citet{Zhao+2016} that assumes a 
constant $T$ of 10~K. In fact, when the same EOS is used, the results 
from the two networks almost overlap with each other, which validates 
the convergence of both chemical networks.

To summarize, the magnetic diffusivities are very sensitive to 
the grain size distribution. In the low density regime (<10$^{10}$~cm$^{-3}$) 
where the diffusion of magnetic fields matters the most for disk formation 
and growth, ambipolar diffusivity $\eta_{\rm AD}$ reaches a maximum 
level for disk formation when the MRN size distribution is truncated 
at 0.1~$\mu$m, and Hall diffusivity $\eta_{\rm Hall}$ reaches a 
maximum level when the distribution is truncated at 0.04~$\mu$m.\footnote{The 
optimal grain size for Hall effect also depends slightly on the 
cosmic-ray ionization rate $\zeta_0^{\rm H_2}$, which changes to 
$a_{\rm min}$$\sim$0.03~$\mu$m for $\zeta_0^{\rm H_2}=5.0\times10^{-17}$~s$^{-1}$
and $a_{\rm min}$$\sim$0.07~$\mu$m for $\zeta_0^{\rm H_2}=1.0\times10^{-18}$~s$^{-1}$ (not shown). 
However, the magnitude of $\eta_{\rm AD}$ roughly scales with 1/$\sqrt{\zeta}$ 
and the optimal grain size of 0.1~$\mu$m stays the same for different 
$\zeta_0^{\rm H_2}$.}
Note that when choosing $a_{\rm min}=0.04~\mu$m, the resulting 
$\eta_{\rm AD}$ is not far from its maximum level with 
$a_{\rm min}=0.1~\mu$m. Nevertheless, the presence 
of VSGs reduces the strength of both ambipolar and Hall diffusivities. 
Furthermore, the Ohmic dissipation only becomes comparable to the other 
two effects at very high densities (a few $\sim$10$^{12}$~cm$^{-3}$), 
similar to the conclusions of other work 
\citep[e.g.,][]{Zhao+2016,Marchand+2016}.

\subsection{Chemical Abundances at High Densities: Tracers for Protostellar Disks}
\label{S.Tracers}

By including the barotropic EOS (Appendix~\ref{App.A}), thermal 
desorption of molecules due to an increased gas temperature is 
simultaneously modeled with the chemical network, which can be used 
to analyze the chemical tracers in protostellar disks. As shown in 
Fig.~\ref{Fig:fld+crd_MRN}, thermal desorption efficiently operates 
for most species at densities $\gtrsim$a few 10$^{11}$~cm$^{-3}$, 
above which neutral-neutral reactions dominate the chemistry. 
At number densities between 10$^{12}$--10$^{13}$~cm$^{-3}$, the main 
gas-phase species are CO, O$_2$, N, N$_2$, CO$_2$, and a small fraction 
of CH$_4$. As temperature rises even higher ($\sim$200--300~K) at 
number densities >10$^{13}$~cm$^{-3}$, H$_2$O start to return to gas-phase 
via thermal desorption, and most O$_2$ are turned into 
CO$_2$ by reacting with CO. But the abundances of CO$_2$, N and CH$_4$ 
continue to decrease towards even higher densities beyond the range 
considered here, leaving only CO, H$_2$O and N$_2$ as the 
most abundant species in early protostellar disks 
\citep[see also, e.g.,][]{Molyarova+2017} based on our network. 
Note that grain surface chemistry is very important 
for the more complex species (e.g. CH$_3$OH and other complex 
organic molecules) which best trace the high density 
and warm regions surrounding the central protostar, including 
protostellar disks; we will describe this more detailed 
chemical work in a future paper.

In comparison, the fractional abundances of ions at high densities 
are more than ten orders of magnitude lower than the major neutral 
species. Our result shows that HCO$^+$ remains as a good ionic tracer 
in the bulk part of the protostellar disk, along with H$_3$O$^+$ which 
is hard to observe. N$_2$H$^+$ should only trace the outer part of 
the disk and show as a ring structure. Note that the ionization 
fraction presented here is regulated by the simple exponential 
attenuation of cosmic-rays; improving the cosmic-ray attenuation 
function or including other sources of ionization 
\citep[e.g., X-ray, thermal ionization; see][]{Armitage2011,Bai2011a}
should provide a more realistic ionization fraction in different parts 
of the protostellar disk. We will explore such topics in an upcoming work.

\section{Discussion \& Summary}
\label{Chap.Summary}

The new chemical network presented here is still under development, 
with a few main ingredients to be implemented, including multiple grain 
charging \citep[e.g.,][]{Ivlev+2016}, grain surface chemistry 
\citep[e.g.,][]{Harada+2017}. 
Multiple charging becomes important for large grains at high temperature 
\citep[e.g.,][]{DraineSutin1987}, and should be investigated in detail 
for our modeling of protoplanetary disks next. 

We noticed that a parallel work by \citet{IqbalWakelam2018} that includes 
grain surface chemistry also uses a similar formulation of cosmic-ray 
desorption rate. However, they assume a constant evaporative cooling 
timescale of 10$^{-5}$~s for all grain sizes, which leads to a 
significant over-estimation of the impact of small grains on the gas-phase 
abundances (cosmic-ray desorption rate increased by many orders of 
magnitude for all species). 
In fact, the chemical abundances computed using the conventional 
pre-factor $f(70~{\rm K})$ is already quite high compared to observations 
\citep[see Fig.~9 of][]{Shen+2004}. 
As stated by \citet{HasegawaHerbst1993}, the evaporative cooling 
timescale is inversely proportional to $k_{\rm des}$, and the 
value of 10$^{-5}$~s is only suitable for 0.1~$\mu$m with 70~K. 
For grains with enough surface sites (>0.03~$\mu$m), this timescale can 
be properly calculated using different $T_{\rm e}$ for the given grain 
size $a$ (see \S~\ref{S.React}). Within the size range [0.03,0.25]~$\mu$m, 
slightly small grains can be cooled 
down by tens of K from the peak temperature ($T_{\rm e}$) by 
evaporating only $\sim$10$^5$ CO molecules, while the temperature 
of larger grains is difficult to reduce, with the reduction of a few K 
requiring the evaporation of a few 10$^6$ CO molecules 
(values obtained by integrating Eq.~\ref{Eq:Cv}). Therefore, 
larger grains stay at their elevated temperatures for a much longer time, 
allowing a sufficient desorption of volatile species from their surfaces. 
In comparison, the peak temperature of small grains is so high that 
the majority of CO and other volatiles evaporate almost instantly, 
the small grain then cools down via sublimation of H$_2$O while 
desorpting other non-volatile species. 
Nonetheless, the evaporation timescale of volatile species 
on small grains should be rather rapid. 

Furthermore, according to the formulation by \citep{HasegawaHerbst1993}, 
the rate of cosmic-ray desorption should have a weak dependence on 
$T_{\rm e}$ for a given species, since both the duty cycle $f(T_{\rm e}(a))$ 
and the desorption rate $k_{\rm des}(i,T_{\rm e}(a))$ in 
Eq.~\ref{Eq:k_crd} have similar exponential dependence on $T_{\rm e}$. 
Adopting a constant evaporating cooling time of 10$^{-5}$~s for 70~K 
alone but much higher $T_{\rm e}(a)$~K in $k_{\rm des}(i,T_{\rm e}(a))$ 
is responsible for the large effect of cosmic-ray desorption 
on the chemical abundances shown in \citet{IqbalWakelam2018}.
Nevertheless, our new formulation of cosmic-ray desorption is 
still based on \citet{HasegawaHerbst1993} and \citet{Acharyya+2011}, 
which is a crude approximation to the stochastic process of 
impinging cosmic-ray particles on grain surfaces.

Despite the simplified chemical model we adopted here, we are able 
to extract a few novel effects of the grain size distribution 
on the desorption of volatile species and on the non-ideal MHD 
diffusivity, along with the key factors operating behind these 
effects: 
\begin{description}
\item(1) The rate of direct desorption of volatile species by 
cosmic-ray from grains with different sizes can vary by many orders of 
magnitude over the range of grain size in the standard MRN size 
distribution, due to the large difference in the elevated temperature 
of grains of different sizes, instead of a constant 70~K 
\citep{HasegawaHerbst1993}. 
\item(2) The desorption of volatile species by cosmic-ray is very 
sensitive to the presence of grains larger than 0.1~$\mu$m, 
because they are heated by cosmic-ray to a lower 
temperature ($\sim$40~K) and the difference in desorption rates among 
volatile species due to the slight difference in binding energies is 
amplified. The resulting chemical abundances show an increased amount of 
nitrogen bearing species in the gas-phase. Particularly, atomic nitrogen 
N is about two orders of magnitude more abundant than CO, while gas-phase 
N$_2$ is a few times more than CO. Accordingly, N$_2$H$^+$ becomes a 
more abundant tracer than HCO$^+$ in dense cores, which provides 
a natural way to explain the well known difference between the 
distribution of CO and N$_2$ related molecules (such as N$_2$H$^+$) 
in dense cores \citep[e.g.,][]{Caselli+2002a,Caselli+2002b}. 
\item(3) Small grains ($\lesssim$few 100~$\AA$) dominate the 
cosmic-ray desorption rate of non-volatile species such OH and H$_2$O. 
However, for such species, direct cosmic-ray desorption is much less 
efficient than the desorption by cosmic-ray induced UV photons.
\item(4) Freeze-out of molecules changes the dominant ion in dense cores 
from the commonly assumed HCO$^+$ to H$_3^+$, yet the relative 
abundance between H$_3^+$ and HCO$^+$ (N$_2$H$^+$) at given densities 
also changes with grain size. The relation of ion densities, if observed, 
may provide a potential way to constrain the grain size in dense cores. 
Besides, the change in the dominant ion species in the gas-phase has 
very limited effect on the magnetic diffusivities. 
\item(5) The magnetic diffusivities are highly dependent on the grain size 
distribution. We confirm the role of VSGs in weakening the efficiency 
of both ambipolar diffusion \citep{Padovani+2014,Zhao+2016} and Hall 
effect with or without freeze-out and desorption. 
At densities below $\sim$10$^{10}$~cm$^{-3}$, the maximum 
ambipolar diffusion is achieved when using a truncated MRN size 
distribution with $a_{\rm min}\sim0.1~\mu$m; and to reach a 
maximum Hall effect, $a_{\rm min}\sim0.04~\mu$m is optimal. 
\end{description}

Our chemical network can be either tabulated to provide 
the magnetic diffusivities for non-ideal MHD simulations or used 
to post-process the simulation results of dense cores and protostellar 
disks. Although much work remains to be done, our results have shown that 
grain evolution in dense molecular clouds has a profound influence 
on both the chemical and dynamical evolution of the system, which 
deserves more self-consistent study in the future.

\section*{Acknowledgements}

We thank Wing-Fai Thi, Seyit H\"{o}c\"{u}k, and Olli Sipil\"{a} 
for inspiring discussions. 
BZ and PC acknowledge support from the European Research Council 
(ERC; project PALs 320620). 
Z.-Y. L. is supported in part by NASA NNX14AB38G and 
NSF AST-1716259.


\appendix

\section{Equation of State}
\label{App.A}

\begin{figure*}
\includegraphics[width=\textwidth]{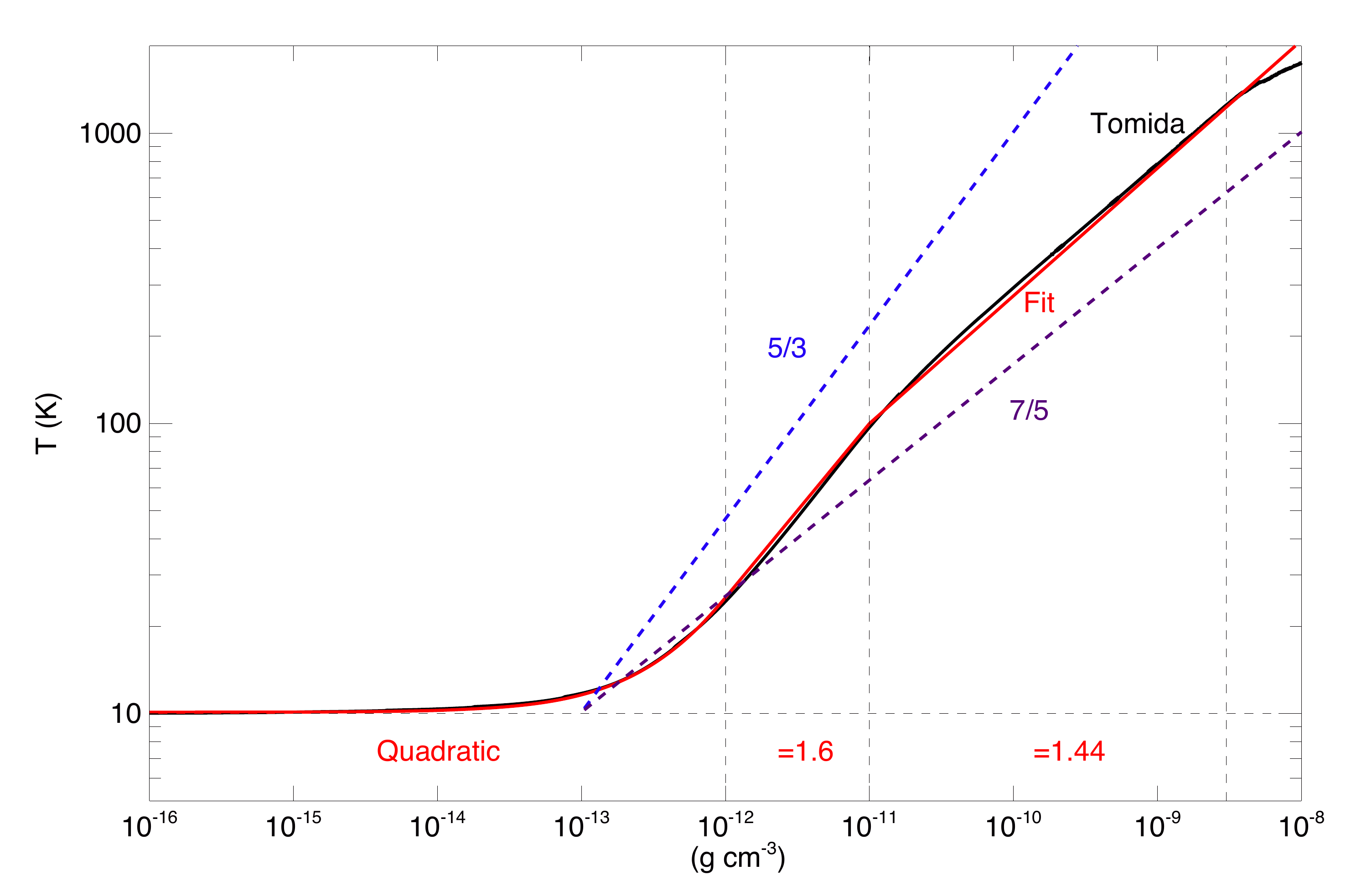}
\caption{Function fitting (red solid curve) for the evolutionary track 
from \citet{Tomida+2013} (black solid curve). The barotropic EOS 
with adiabatic indices of 5/3 and 7/5 are shown in blue dashed and 
purple dashed lines.}
\label{Fig:EOS}
\end{figure*}

We use a broken power law profile for the equation of state (EOS), 
fitted to mimic the radiative transfer results of \citet{Tomida+2013}: 
\begin{equation}
T= \left\{\begin{array}{lcl} T_0 + 1.5{\rho \over 10^{-13}} & \mbox{for} & \rho<10^{-12} \\
(T_0+15)({\rho \over 10^{-12}})^{0.6} & \mbox{for} & 10^{-12} \leqslant \rho \leqslant 10^{-11} \\
10^{0.6}(T_0+15)({\rho \over 10^{-11}})^{0.44} & \mbox{for} & 10^{-11} \leqslant \rho \leqslant 3\times10^{-9} \end{array}\right.
\end{equation}
where T$_0 = 10$~K. The comparison for different EOS is shown in 
Fig.~\ref{Fig:EOS}. The same EOS is adopted in the study of disk formation 
by \citet{Zhao+2018}. 

\section{Magnetic Diffusivity and Conductivity}
\label{App.B}

The three non-ideal MHD coefficients can be expressed in terms of 
the components of the conductivity tensor $\sigma$ 
\citep[e.g.,][]{Wardle2007}: 
\begin{eqnarray}
\eta_{\rm AD} & = & {c^2 \over 4\pi} ({\sigma_{\rm P} \over \sigma_{\rm P}^2 + \sigma_{\rm H}^2} - {1\over \sigma_{\parallel}})~, \label{Eq:MHDcoef1}\\
\eta_{\rm Ohm} & = & {c^2 \over {4\pi \sigma_{\parallel}}}~, \label{Eq:MHDcoef2}\\
\eta_{\rm Hall} & = & {c^2 \over 4\pi} ({\sigma_{\rm H} \over \sigma_{\rm P}^2 + \sigma_{\rm H}^2})~; \label{Eq:MHDcoef3}
\end{eqnarray}
where the parallel $\sigma_{\parallel}$, Pedersen $\sigma_{\rm P}$, 
and Hall $\sigma_{\rm H}$ conductivities are related to the 
Hall parameter $\beta_{i,\rm H_2}$ as: 
\begin{eqnarray}
\sigma_{\parallel} & = & {{e c n({\rm H}_2)} \over B} \sum_i Z_i x_i \beta_{i,\rm H_2}~, \label{Eq:conduct1}\\
\sigma_{\rm P} & = & {{e c n({\rm H}_2)} \over B} \sum_i {{Z_i x_i \beta_{i, \rm H_2}} \over {1+\beta_{i,\rm H_2}^2}}~,\label{Eq:conduct2}\\
\sigma_{\rm H} & = & {{e c n({\rm H}_2)} \over B} \sum_i {{Z_i x_i} \over {1+\beta_{i,\rm H_2}^2}}~; \label{Eq:conduct3}
\end{eqnarray}
where $x_i$ is the abundance of charged species $i$ with respect to 
H$_2$ molecules.The Hall parameter $\beta_{i,\rm H_2}$ is the key quantity 
that determines the relative importance of the Lorentz and drag forces 
for each charged species $i$ in a sea of neutral H$_2$ molecules. 
It is defined as: 
\begin{equation}
\label{Eq:mtr}
\beta_{i,\rm H_2} = ({{Z_i e B} \over {m_i c}}) {{m_i+m_{\rm H_2}} \over {\mu m_{\rm H} n({\rm H}_2) <\sigma v>_{i,\rm H_2}}}~,
\end{equation}
where $m_i$ and $Z_i e$ are the mass and the charge of charged species i, 
respectively, 
and <$\sigma v$>$_{\rm i,H_2}$ is the momentum transfer rate coefficient, 
parametrized as a function of temperature \citep{PintoGalli2008}, 
which quantifies the collisional coupling between neutral (H$_2$) and 
charged ($i$) species. 

\section{Chemical Abundances for 
\texorpdfstring{$\lowercase{a_{\rm min}=0.04~\mu$m}}{} Case}
\label{App.C}

\begin{figure*}
\includegraphics[width=\textwidth]{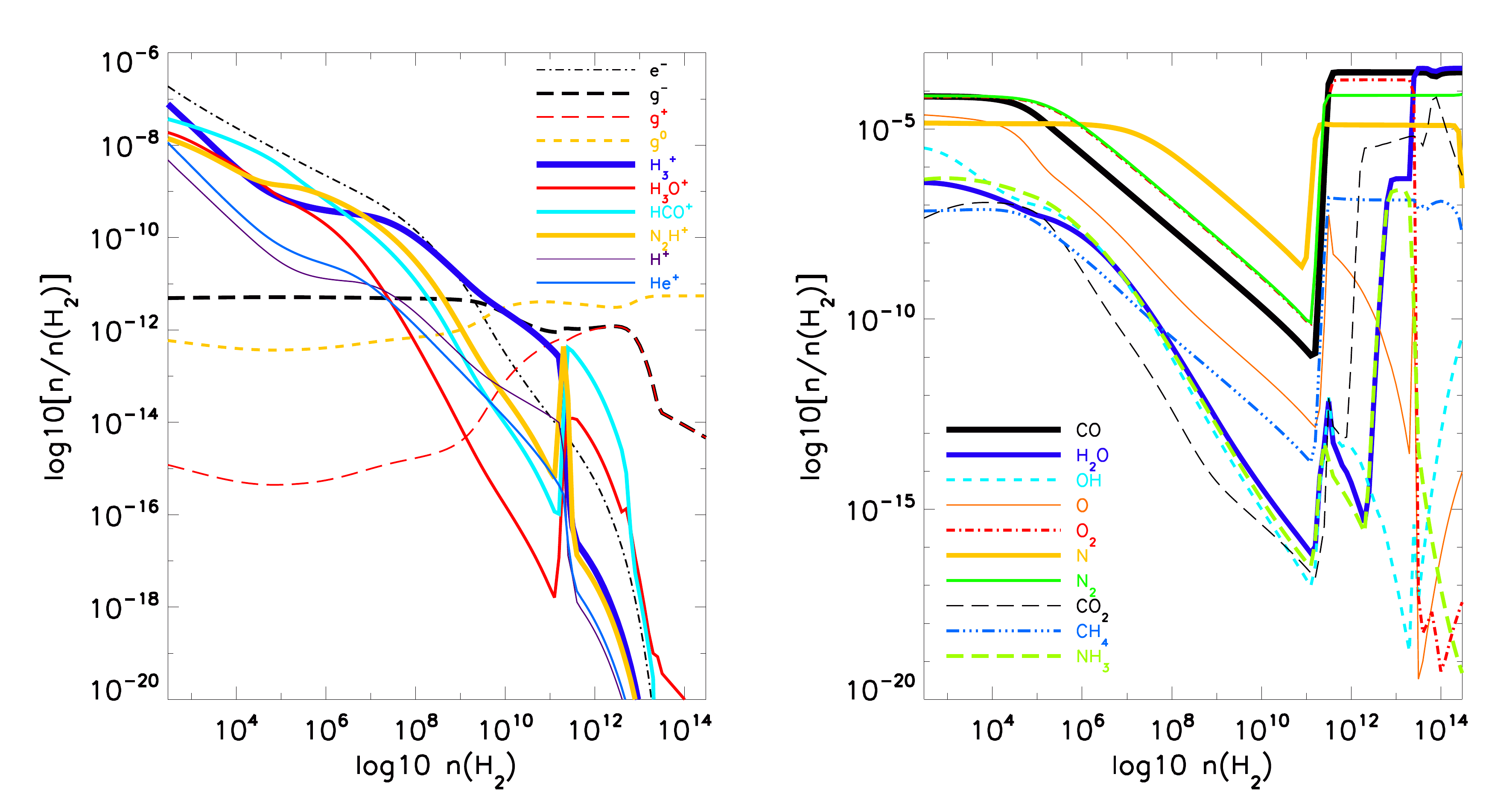}
\caption{Fractional abundances of the main ion and neutral species 
computed using the new formulation of cosmic-ray desorption rate for 
$a_{\rm min}=0.04~\mu$m.}
\label{Fig:fld+crd_trMRN400}
\end{figure*}

Fig.~\ref{Fig:fld+crd_trMRN400} shows the fractional abundances computed 
using a truncated MRN size distribution with $a_{\rm min}=0.04~\mu$m, 
which is used for the discussion in \S~\ref{S.Freeze}. 


\bsp	
\label{lastpage}
\end{document}